\documentclass[aps,pre,preprint,superscriptaddress]{revtex4-1}
\usepackage{graphicx}
\usepackage{dcolumn}
\usepackage[centertags]{amsmath}
\usepackage{tikz,mathpazo}
\usetikzlibrary{shapes.geometric, arrows}
\usepackage{flowchart}
\usepackage{caption}

\usepackage{amsfonts}
\usepackage{amssymb}
\usepackage{amsthm}
\usepackage{newlfont}
\usepackage{color}
\usepackage{natbib}
\usepackage{subfigure}
\usepackage{float}

\DeclareCaptionLabelFormat{cont}{#1~#2\alph{ContinuedFloat}}
\begin{document}

\title{Two-Phase Dynamics of DNA Supercoiling  based on DNA Polymer Physics}

\textrm{\textrm{\emph{}}}

\author{Biao Wan}
\address{Beijing Computational Science Research Center, Beijing 100193,China}

\author{Jin Yu}
\email{jin.yu@uci.edu}
\address{Department of Physics and Astronomy, Department of Chemistry, NSF-Simons Center for Multiscale Cell Fate Research, University of California, Irvine, CA 92697, USA}

\begin{abstract}

DNA supercoils are generated in genome regulation processes such as transcription and replication, and provide mechanical feedback to such processes. Under tension, DNA supercoil can present a coexistence state of plectonemic (P) and stretched (S) phases. Experiments have revealed  the dynamic behaviors of plectoneme, e.g. diffusion, nucleation and hopping.  To represent these dynamics with computational changes, we demonstrated  first the fast  dynamics on the DNA to reach torque equilibrium within the P and S phases,  and then identified the two-phase boundaries as collective slow variables to describe the essential dynamics. According to the time scale separation demonstrated here, we accordingly developed a two-phase model on the dynamics  of DNA supercoiling, which can capture physiologically relevant events across time scales of several orders of magnitudes.  In this model, we systematically characterized  the slow dynamics  between the two phases, and compared the numerical results  with that from the DNA polymer physics-based  worm-like chain model. The supercoiling dynamics, including the nucleation, diffusion, and hopping of plectoneme, have been well represented and reproduced, using the two-phase dynamic model, at trivial computational costs. Our current developments, therefore, can be implemented to explore multi-scale physical mechanisms of the DNA supercoiling-dependent physiological  processes.

\end{abstract}

\maketitle{}

\section*{INTRODUCTION}

Supercoiling is ubiquitous in cellular DNA, which results from the topology of double-helical DNA \cite{Bates1993DNA,Fuller1971The,Koster2010Cellular,Lavelle2014Pack}. For eukaryotic DNA, the large structural domains ($\sim 1$ M$b$) emerge in the genome as the topologically associating domains (TADs)\cite{dixon2012topological,Glinsky2015Rapidly,2019Transcriptions}, which are further divided into smaller supercoiling domains($\sim10^5 bp$)\cite{2013Transcription,2014Supercoiling,2016Effects}.  For the prokaryotic  DNA, comparatively  large structural domains  and  smaller supercoiling domains also present (e.g., $E$. $coli$ with  an average domain size, $\sim10^{4}$ $bp$) \cite{Postow2004Topological,2014Supercoiling}. Supercoils are generated in many crucial genetic processes  and in turn regulates corresponding processes, including transcription, replication  and  genome condensation\cite{Liu1987Supercoiling,Ma2013Transcription,Chong2014Mechanism,Bates1993DNA,burnier2008dna,Schvartzman2013The,Hirano2000Chromosome,strick2004real}. During transcription elongation, a RNA polymerase (RNAP) moves along the helical DNA and generates  twists, which subsequently create positive supercoiling ahead of the RNAP or downstream along the transcription direction and negative supercoiling behind or upstream\cite{Liu1987Supercoiling}.  The accumulation of the torsional stress can slow down or even stall the transcription elongation\cite{Ma2013Transcription,Chong2014Mechanism}.   
Recent experiments show  long-distance cooperative and antagonistic dynamics of multiple RNAPs via supercoiling\cite{kim2019long}, which justify the role of supercoiling as a  long-range mediator\cite{Teves2014DNA}.  A previous modeling study pointed out that multiple RNAPs facilitate elongation by reducing collisions or traffic jam by torques\cite{Tantale2016A}; and another model suggested that multiple RNAPs also induce strong supercoiling to bring significant transcription fluctuations\cite{Jing2018How}.  
In addition, supercoiling contributes to the  chromosome condensation\cite{Hirano2000Chromosome,strick2004real}, during which the negative supercoils accumulate  in the protein-free region of the DNA \cite{kimura1997atp}.  Under tension, the accumulative supercoiling presents  twisted and interwound  coils called plectonemes, which may act as a driving force for compacting the chromatin fiber\cite{kimura1997atp,Swedlow2003The}.

Supercoil can be partitioned into two parts, twist  and writhe,  which play distinct roles in genomic processes\cite{Seol2016The}. Underwound DNA (with negative twists)  can melt the  secondary structures of DNA that can attract  regulatory proteins\cite{Ma2014Interplay,Ma2014RNA,Seol2016The}. In contrast, overwound DNA (with positive twists) can hinder enzymatic activities associated with opening of the DNA duplex, such as in transcription initiation and elongation by RNAP\cite{Ma2013Transcription,Chong2014Mechanism}. On the other hand, writhe characterizes spatial  crossings of DNA segments\cite{Fuller1971The,Fuller1978Decomposition}, accordingly facilitating distant  interactions on DNA\cite{Parker1991Dynamics,liu2001dna,Huang2001Dynamics,polikanov2007probability}. Writhe can be in a solenoidal or a plectonemic form, or a combination of both. Under an externally stretching force, supercoiled DNA can present a coexistence state of both\cite{Marko1997Supercoiled,strick1999phase,Ghatak2005Solenoids,Marko2007Torque}, in which the plectonemic coils form the plectonemic (P) phase, while the  solenoidal coils are almost straightly stretched, forming the stretched (S) phase\cite{Marko2007Torque,forte2019plectoneme}.

DNA supercoiling has been quantitatively studied with single-molecule techniques\cite{marko1994fluctuations,Bustamante1994Entropic,Strick1996The,strick1998behavior,strick1999phase,bustamante2000single,strick2000twisting,Koster2005Friction,van2012Dynamics,Bustamante2016Ten}. The extension-twist curves from experiments reveal a coupling between supercoiling (linking number) and DNA  extension under stretching force\cite{marko1994fluctuations,Bustamante1994Entropic,Strick1996The}. The discontinuities on torque and extension versus linking number were also observed in experiment\cite{Forth2008Abrupt}. Experiments with a  magnetic tweezers pulling a fluorescently labeled  DNA have directly imaged the coexistence phase, and remarkably, measured the time-dependent supercoiling dynamics, e.g., plectoneme diffusion, nucleation and hopping\cite{van2012Dynamics}.  The propagation of plectoneme along DNA  via diffusion presents conformational rearrangement of DNA at about hundreds bps per second.  The hopping happens comparatively fast  and a fastest one observed  takes tens of milliseconds for a displacement of thousands of bps\cite{van2012Dynamics}.   The surviving time of the individual plectoneme from nucleation to vanishing, i.e., the plectoneme life-time, spans  from milliseconds  to several seconds.

 In accompany with the experiments, theoretical  studies of DNA supercoiling have been developed based on the worm-like chain (WLC) model\cite{Marko1995Stretching,Marko1995Statistical,Moroz1997Torsional,Marko2007Torque,Phillips2009Physical}.    The statistical mechanics of the WLC model of DNA provides a description of the coupling between supercoiling and DNA extension\cite{Moroz1997Torsional,Marko1998DNA},  and reveals the  pre-plectonemic loops formed by bending DNA, i.e., buckling transition\cite{Daniels2011Nucleation}, which  give rise to the discontinuities on torque and extension versus linking number curves\cite{Marko2012Competition,Emanuel2013Multiplectoneme}. Further, studies  show that under a stretching  force, the coexistence  of the S and P phases is maintained under a  coexistence torque that is a function of the stretching force\cite{Marko2007Torque}. 

 Nevertheless, the related analytic theories concern mostly  the equilibrium and static behaviors, without elucidating the time-dependent dynamic processes.    The experiments suggest that the dynamics of plectonemes essentially covers multiple orders of timescales\cite{van2012Dynamics}.   In this study, we focus  on building  a model to describe these dynamic processes across the time scales.
  
Apart from the analytic theories, computational simulations and numerical methods have offered effective ways to probe supercoiling dynamics. Atomistic molecular dynamics (MD) simulations offer the finest details  of such dynamics. For example, MD simulations of DNA mini-circles ($10^{2}$ $b$) reveal  writhe fluctuation and configuration diversity\cite{Mitchell2013Thermodynamics,Mitchell2011Atomistic,R2015Structural}. To improve computational efficiency, a coarse-grained model that treats nucleotides as beads with three interaction sites,  called oxDNA, was developed\cite{matek2015plectoneme}. Furthermore, by unitizing MD and the coarse-grained simulations, the  multi-scale dynamics of supercoiling have been studied\cite{sutthibutpong2016long,hirsh2013structural}. Nevertheless, computational costs of these models  are high and the simulation time scales are limited by microseconds\cite{2016Multiscale}.    The models without considering DNA sequence structure  are capable of exceeding such a limit.  An elasto-dynamic model (i.e., the Kircholff rod), for example, has been developed, by which  DNA is characterized in terms of  a continuum rod while omitting thermal fluctuations\cite{Goyal2005Nonlinear}. The corresponding applications were carried out  on supercoil removal\cite{lillian2011a}, and on  compressed  DNA inside  bacteriophage cavity to allow DNA ejection\cite{Hirsh2012A}.  In particular, for a representative polymer physics model, the numerical approach of the WLC model, i.e., the discrete worm-like chain (dWLC) method, has been widely implemented in simulating DNA with the thermal fluctuations incorporated\cite{Vologodskii2006Simulation}.   For example, the Monte Carlo (MC) simulations of the WLC model have been used in  describing DNA thermodynamics, conformation  and site juxtaposition\cite{Alexander1992Conformational,Klenin1991Computer,polikanov2007probability}. The Brownian dynamics (BD) simulations, on the other hand, have been applied more widely in studying  supercoiling\cite{Vologodskii2006Simulation,Klenin1998A}. Such type of studies successfully show the  plectoneme diffusion\cite{Bell2012Simulation},  supercoiling conformations \cite{Alexander1992Conformational,babamohammadi2020traveling}, buckling transition\cite{Daniels2011Nucleation,Walker2018Dynamics,ott2020dynamics}, modeling a DNA wrapped around a model histone\cite{T2009Brownian}, and supercoil removal by nicking\cite{Ivenso2016Simulation}. Besides, other semi-flexible polymer models with Lennard-Jones potential have also been utilized to investigate the mechanism for  supercoiling  by rotating the ends\cite{wada2009plectoneme}, or using interwoven and braided polymers to produce plectonemes\cite{forte2019plectoneme}. 
 
  The computational costs of the above models are still quite significant for studying the plectoneme dynamics. For example, by employing the dWLC method, simulating the plectoneme dynamics over a second requires weeks of CPU time\cite{Ivenso2016Simulation,ott2020dynamics}.     In this study, we developed a  two-phase (S and P phases) dynamic model of DNA supercoiling at trivial computational cost based on the WLC polymer physics model of DNA.   In solvent, supercoiled DNA experiences frictional forces on the two phases and the energy gradient that drives the transformation between the S and P phases.   Thus the first thing is to identify the fast dynamics comparing to the phase-transformation  and possibly integratae them out to reduce the degrees of freedom\cite{2004Adaptive,zhang2010enhanced}.  Here we  demonstrate first the dynamics and timescales of the torque transport within S and P phases that define the fast dynamics. Then we choose  interphase boundaries as the slow dynamic observables and obtain the  associated energetics.  Subsequently, we derive the Langevin dynamics of the slow observables.  As a result, this  implies that DNA supercoil can fast propagate as twist (torque) and slowly propagate as writhe (or plectonemes) along DNA.   For calibration and consistency check, we compare the numerical results of the two-phase dynamic model with those from the dWLC method. For applications, we reproduce the supercoil dynamics on  plectoneme diffusion, nucleation and vanishing as being measured experimentally, as well as the force/ionic strength-dependent dynamics.  Finally, plectoneme hopping and the associated energetics transformations between the two phases are monitored.

\section*{RESULTS}
 Here, we study dynamics of the DNA supercoiling and build a two-phase dynamic model with a focus  on DNA of $10^3$ to $10^4$ $b$. Such sizes of DNAs are widely used in the single-molecule experiments\cite{Crut2007Fast,van2012Dynamics,Ma2013Transcription}.  DNAs  shorter than  several hundreds bps hardly bend to  form a stable plectonmic helix;  for DNAs  longer than $10^{4}$ $b$, however, the supercoiling domain may emerge and take over\cite{dixon2012topological,Glinsky2015Rapidly}.

\subsection*{Fast torsional equilibrium on the stretched (S) and plectonemic (P) phases}
  
  In this section, we describe the dynamics of torque transport  reaching to equilibrium within the S and P phases,  and show that such dynamics are fast comparing to the transformation between the two phases.
  
  \subsubsection*{Torque transport on stretched DNA or S phase}
  
 We study the dynamics of  torque propagation on supercoiled DNA based on the WLC model numerically, i.e., via an implementation of the dWLC method (see Text S1 in the Supporting Material).

In the WLC model, harmonic potentials are used in representing the elasticities of DNA\cite{Bates1993DNA,Marko1995Statistical,Klenin1998A}.
    The twist energy density is defined as 
    \begin{equation}
    \mathcal{E}_{twist} = \frac{1}{2}k_{B}Tl_t(\frac{\Delta \theta}{\Delta u})^2\tag{A1}
    \label{eq:a1}
    \end{equation}
     where $k_B$ is the Boltzmann constant, $T$ is the room temperature, $l_t$ is the torsional persistence length, i.e., $l_t=75 nm$,  $\Delta \theta$ is  the twist on the segment $\Delta u$.
   We introduce an internal torque $T_q^{in}$ as a measure of torsional stress on DNA, $T_q^{in}\equiv k_{B}Tl_t\frac{\Delta \theta}{\Delta u}$ (Torque is proportional to twist).    Based on the elastic properties of DNA, the torque transport  along DNA can be written as (see  Text S2), 
  \begin{equation}  
\frac{\partial T_q^{in}}{\partial t}=\frac{l_tk_{B}T}{\zeta_{R}}\frac{\partial ^{2}T_q^{in}}{\partial u^{2}}\tag{A2}
\label{eq:a2}
\end{equation}
where drag coefficient   for a rotational cylinder $\zeta_{R}=4\pi R^{2}\eta$, and the radius of DNA $R=2 nm$ and the viscosity $\eta=0.001 kg/m/s)$. A solution to Eq \ref{eq:a2} can be seen in Text S2.  The characteristic time for torque equilibrium time can be reached at  $ 2.5\frac{\zeta_{R}}{l_tk_{B}T}L^2$(about 0.1$ms$ for $3000 bp$) (Text S2), independent of stretching force, torque and linking number.  For example, a twist pulse starting at one end of  the stretched DNA or S phase will spread over  $10^{3}$ $bp$ within $10^{-2} ms$ and  $10^{4}$ $bp$ within $1 ms$.   Compared with the dynamics of plectoneme of $\sim10^{3}$ $bp$ that happens at the time scale from $ms$ to $s$, e.g. diffusion and hopping, the torque transport within the S phase is fast.

  \begin{figure}[H]
  \centering
    \subfigure[]{
    \label{fig:1:a}
    \includegraphics[width=0.28\textwidth,angle=0]{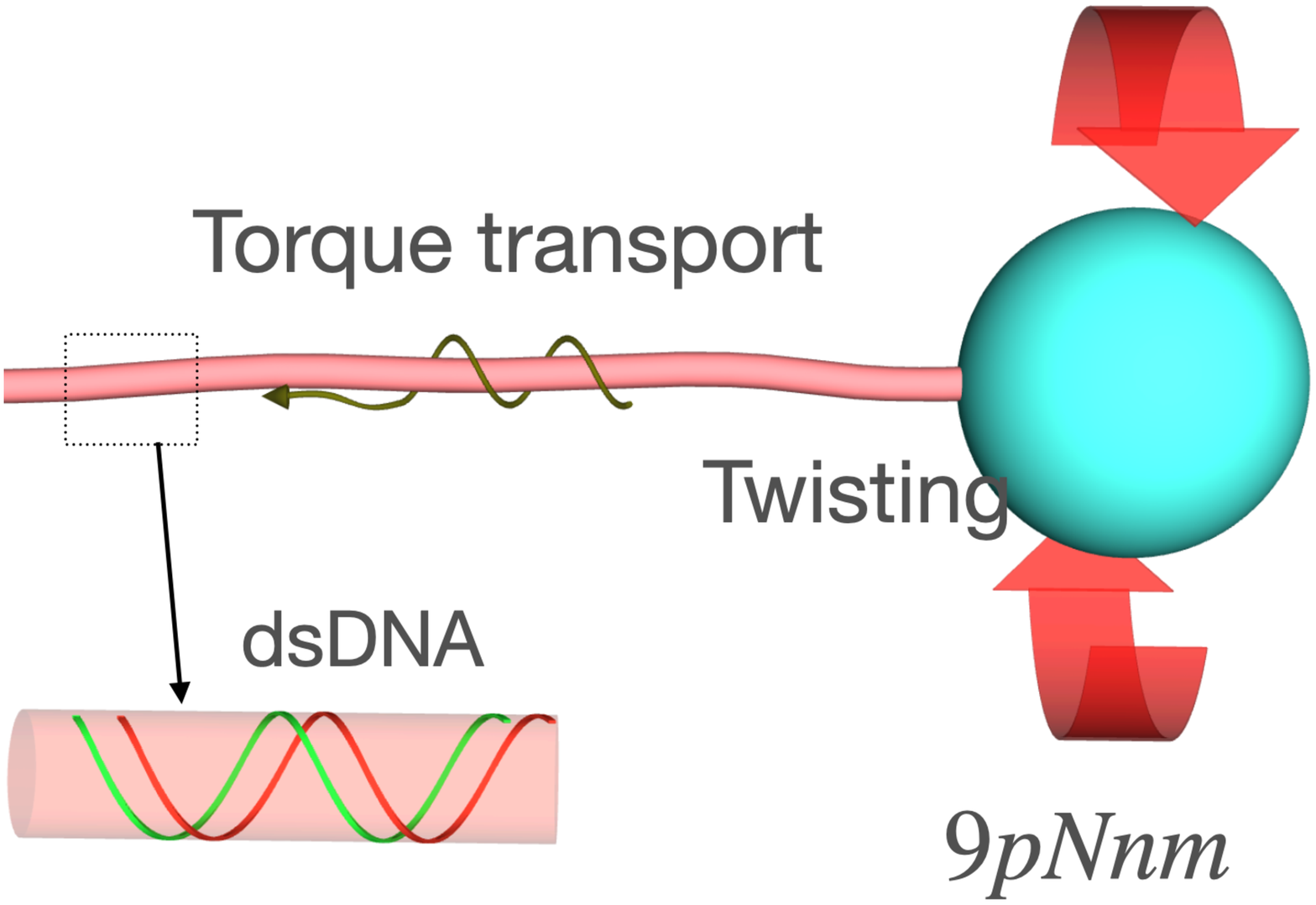}}
    \subfigure[]{
    \label{fig:1:b}
    \includegraphics[width=0.33\textwidth,angle=0]{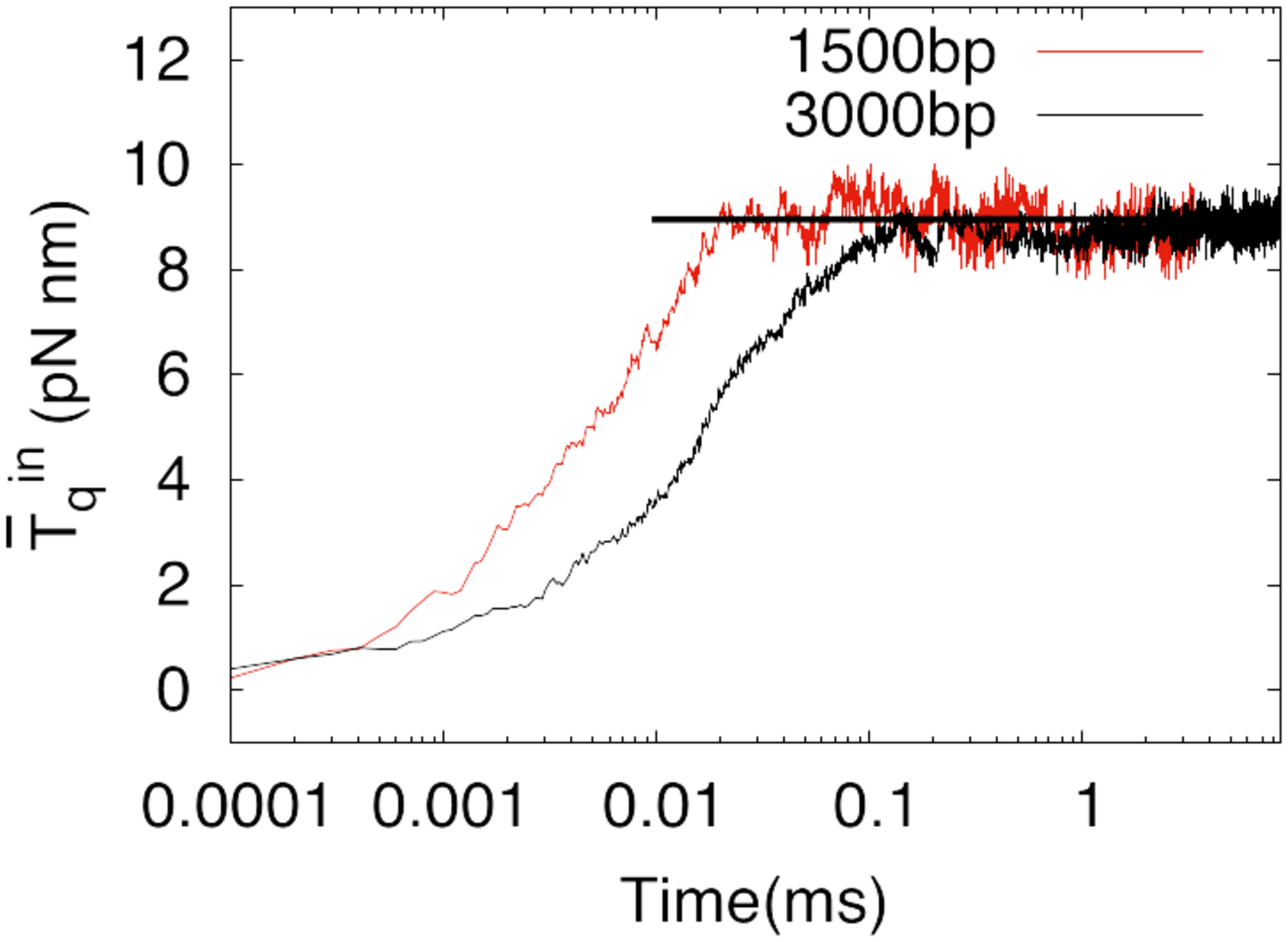}}
        \subfigure[]{
    \label{fig:1:c}
    \includegraphics[width=0.33\textwidth,angle=0]{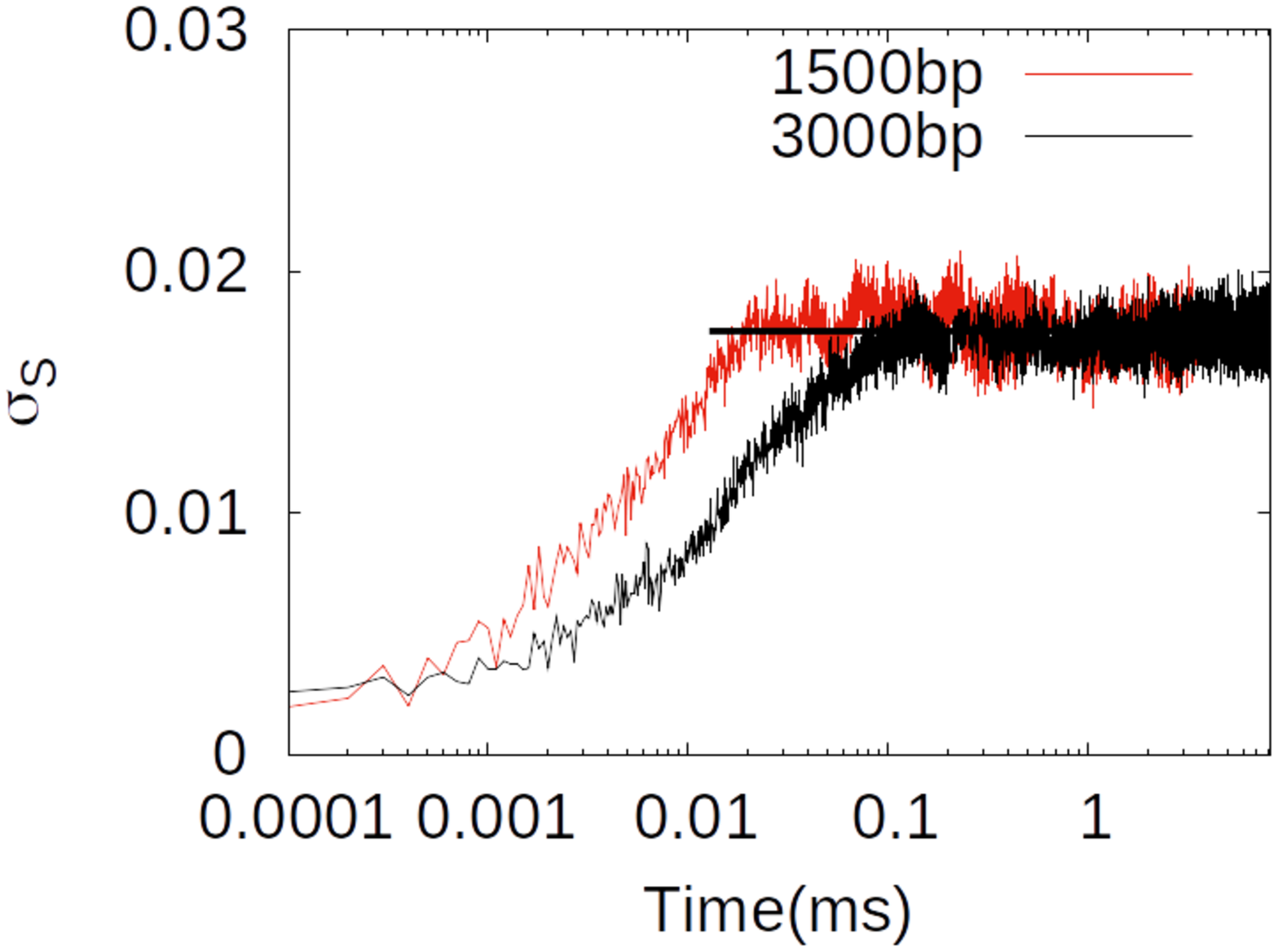}}
    \subfigure[]{
    \label{fig:1:d}
    \includegraphics[width=0.28\textwidth,angle=0]{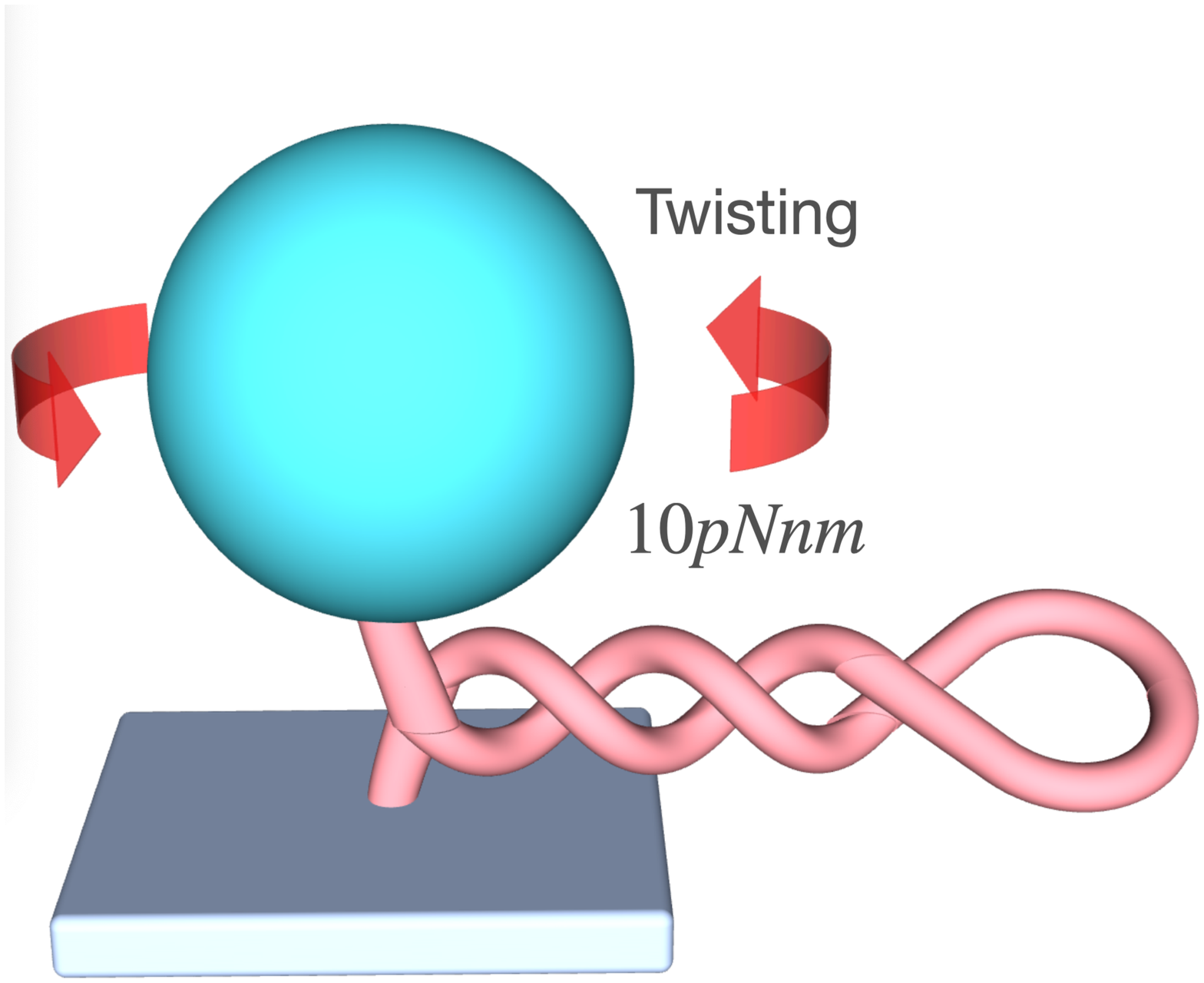}}   
        \subfigure[]{
    \label{fig:1:e}
    \includegraphics[width=0.33\textwidth,angle=0]{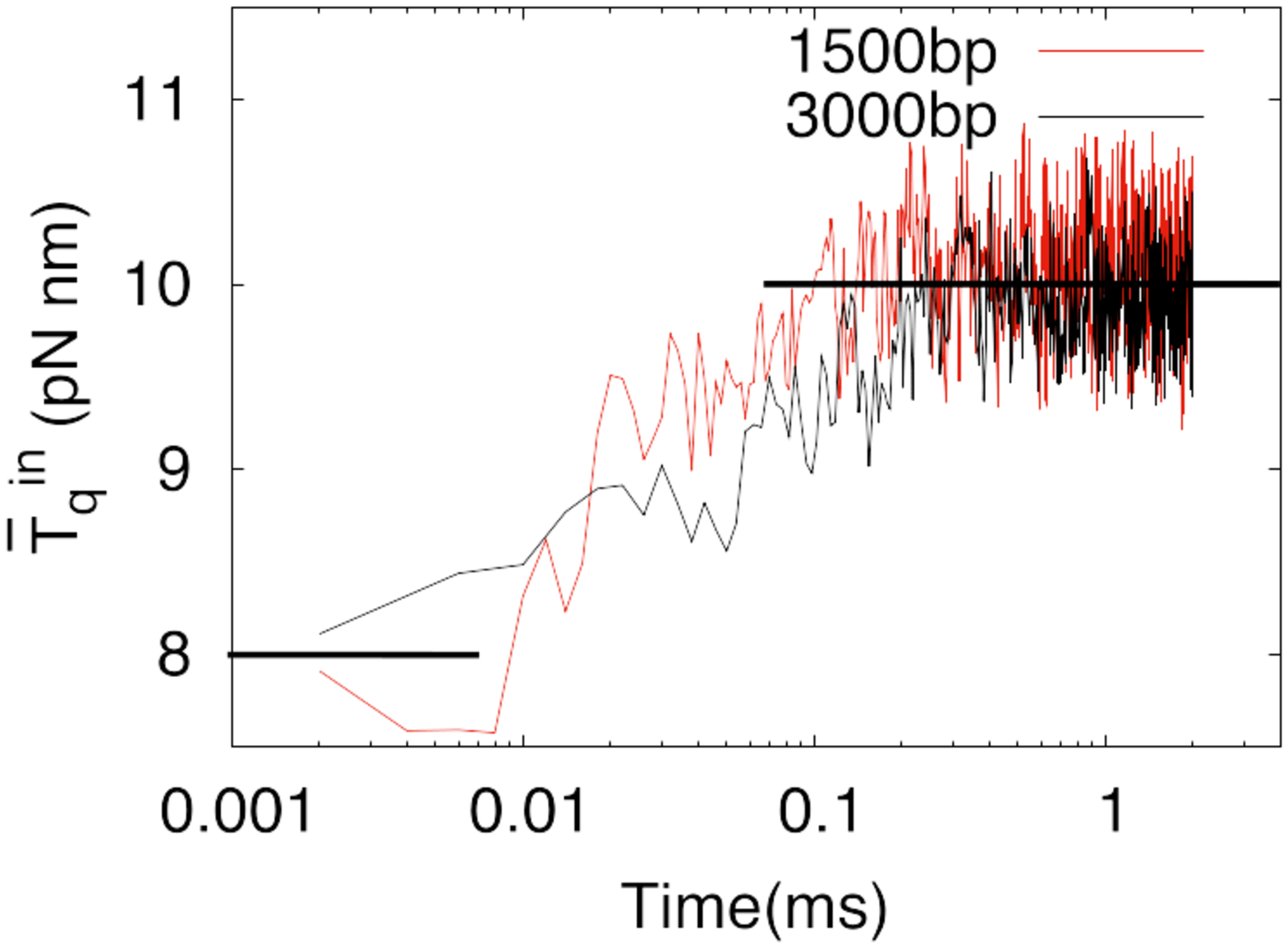}}
    \subfigure[]{
    \label{fig:1:f}
    \includegraphics[width=0.33\textwidth,angle=0]{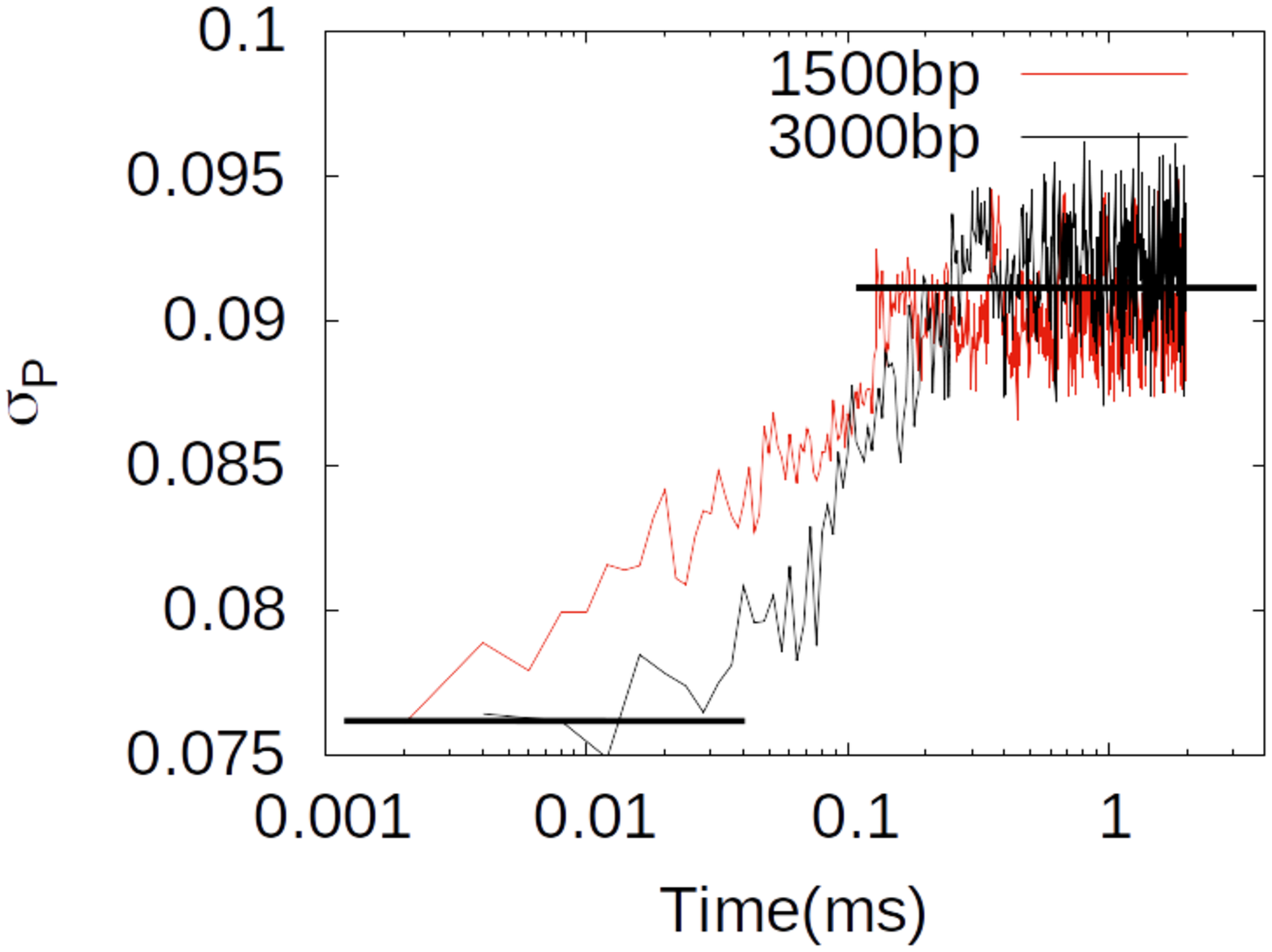}}
         
    \caption{\small Measuring the torque transport on the stretched (S) phase and the plectonemic (P) phase of the supercoiled DNA based on the dWLC, respectively. (a)The schematic shows the torque transport along the S phase following an external torque $T_q^{ex}$ promptly  applied from 0 $pN nm$ to 9 $pN nm$ at $t=0$.   (b) The internal torque of DNA at length of $1500 bp$ and $3000 bp$,  $\bar{T}_q^{in}$ rises to equilibrium (9 $pN nm$) within 0.1 $ms$ to follow $T_q^{ex}$.   (c) The excess linking number density $\sigma_S$   reaches to equilibrium $\sigma_S^{eq}=0.17$ within 0.1$ms$. (d) The schematic shows that  a pure plectonemic DNA follows an promptly applied torque,$T_q^{ex}$ from an initial $8 pN nm$ to maintains the plectonemic helix ($t<0$) to $10 pN nm$ at $t=0$. The internal torque along DNA  consequently propagates on plectoneme. (e) The mean torque of DNA at length of  $1500 bp$ and $3000 bp$, $\bar{T}_q^{in}$ reaches to equilibrium (10 $pN nm$) within $0.3 ms$. (f) The excess linking number density $\sigma_P$  reaches  to  equilibrium value $\sigma_P^{eq}=0.091$ within 0.3 $ms$. }
  \label{fig:1}
  \end{figure}

   Fig.\ref{fig:1:a} shows the scheme for simulating the torque transport on a stretched DNA using the dWLC method.   Under a constant stretching force $f=0.4$ $pN$,  DNA is twisted by an external torque $T_q^{ex}=9pN nm$ starting at $t=0$ ($T_q^{ex}=0$ at $t<0$).   After that, the torque is kept constant and the torsional response of DNA is monitored.
      
   The measurements of the mean internal torque of DNA  at lengths of 1500 $bp$ and 3000 $bp$ are shown  in Fig. \ref{fig:1:b}. After being initially twisted, the mean internal torque, $\bar{T}_{q}^{in}$, reaches  to equilibrium (9 $pN nm$) at  about $0.03 ms$ for 1500 $bp$ and about $0.1$ $ms$ for 3000 $bp$, consistent with the torque equilibrium time $2.5\frac{\zeta_{R}}{l_tk_{B}T}L^2$.
   
  Supercoiling can be quantified by the excess linking number $\Delta Lk$ or the linking number density $\sigma$ (see Text S3). The free energy density  of a stretched DNA is, $\mathcal{F}_{S}=\frac{1}{2}c_S(\sigma_S^{eq})^2+\mathcal{F}_0$(Text S4), including the  contributions from torsion and force stretching\cite{Marko1998DNA,Marko2007Torque}, where $c_S$ stands for the torsional stiffness, which is a function of $f$, $l_t$ and $l_b$ ($l_b=50nm$, the DNA persistence length), $\sigma_S^{eq}$ is the equilibrium linking number density of the stretched phase,  and$\mathcal{F}_0$ is the free energy density without torsion, representing the stretching of the WLC\cite{Marko1995Stretching}.     At equilibrium, the  torque is defined as $\frac{1}{\Omega_0}\frac{\partial\mathcal{F}_{S}}{\partial \sigma_S^{eq}}$\cite{Marko2007Torque}, where the constant $\Omega_0=1.85nm^{-1}$ is the rotation angle of the DNA backbone per unit length\cite{Bates1993DNA,Marko2007Torque}. Thus, we have
   \begin{equation}
   \sigma_S^{eq}=\frac{\Omega_0\bar{T}_q^{in}}{c_S} \tag{A3}
   \label{eq:a3}
     \end{equation} 
  We  accordingly measure  the excess linking number density of the stretched DNA  as shown in Fig. \ref{fig:1:c}. $\sigma_S$ reaches to the equilibrium value  at  0.03$ms$ for 1500 $bp$ and 0.1 $ms$ for 3000 $bp$.     
   
\subsubsection*{Torque transport on plectoneme}
  
 A plectoneme can be  considered as a pair of  rods anti-parallel interwound around each other under an external torque\cite{1997The,Marko2007Torque}.      Thus, an empirical form of the energy of the plectoneme\cite{Klenin1991Computer,1997The,Marko2007Torque} can be utilized,  
  
    \begin{equation}
      \mathcal{F}_P=\frac{1}{2}c_P\sigma_P^2,\tag{A4}
      \label{eq:a4}      
    \end{equation}
   where $c_P\equiv l_pk_BT\Omega_0^2$ is the torsional stiffness of plectoneme, and $l_p$ is the torsional persistence length of the plectonemic helix, depending on the ionic strength and often taking value of 21 to 27 $nm$ \cite{1997The}, and $\sigma_P$ is the excess linking number density of the plectonemes.  The internal toque is defined similarly as that of the stretched DNA, i.e., $T_q^{in}= k_{B}Tl_p\sigma_P$. Nevertheless, torque transport on plectoneme can not be described as the twist angle $\theta$ propagation on DNA.   That is because that the change of  twist of the plectomeical helix also influences  the writhe, and in turn the writhe change feeds back to the twist  propagation.   We  introduce a plectonemic angle $\Theta$  (see Fig. S1), similar to $\theta$, as $T_q^{in}= k_{B}Tl_p\frac{\Delta \Theta}{\Delta u}$. Thus we consider the torque transport on plectoneme as the propagation of the  plectonemical  angle, i.e.,
\begin{equation} 
\frac{\partial T_q^{in}}{\partial t}=\frac{l_pk_{B}T}{\zeta_{P}}\frac{\partial ^{2} T_q^{in}}{\partial u^{2}},\tag{A5}
\label{eq:a5}
\end{equation} 
where $\zeta_{P}\approx2\pi R_{plect}^{2}\eta$ is the drag coefficient due to motions of the two parallel rods and $R_{plect}$ the radius of plectonemes\cite{Neukirch2011Analytical,van2012Dynamics}, about 2.5 nm. The time for torque equilibrium on plectoneme with contour length $L$ can be calculated as $ 2.5\frac{\zeta_{P}}{l_pk_{B}T}\frac{L^2}{4}$, about $2.6$ times comparing to that on the stretched DNA. i.e, about  0.26 $ms$ for torque equilibrium for 3000 $bp$.

  Fig. \ref{fig:1:d} illustrates the setup of torque transport on a pure plectoneme, which is originally maintained by a pre-existing external torque $T_q^{ex}=8pN nm$($t<0$). Then the external torque jumps to $T_q^{ex}=10 pN nm$ at $t\geqslant 0$, and the ensuing torsional response of the plectoneme is monitored. 
    
  The measurements of the internal torque of plectonemes at lengths of 1500 $bp$ and 3000 $bp$ are shown  in Fig. \ref{fig:1:e}. Following the torque jumping at $t=0$, the  mean internal torques, $\bar{T}_{q}^{in}$ rises from $8 pN nm$ to $10 pN nm$ at about 0.1 and 0.3 $ms$ for 1500 $bp$ and 3000$bp$, respectively, as estimated above.  

In equilibrium, Eq. \ref{eq:a4} becomes $\mathcal{F}_P=\frac{1}{2}c_P(\sigma_P^{eq})^2$. The equilibrium torque is defined as $\frac{1}{\Omega_0}\frac{\partial\mathcal{F}_{P}}{\partial \sigma_P^{eq}}$. Thus    
 \begin{equation}
   \sigma_P^{eq}=\frac{\Omega_0\bar{T}_q^{in}}{c_P} \tag{A6}
   \label{eq:a6}
     \end{equation}   
  We also numerically measure  the  excess linking number density of the plectoneme as shown  in Fig. \ref{fig:1:f}.   As Eq \ref{eq:a6} suggests, $\sigma_P^{eq}=\frac{T_q^{ex}\Omega_0}{c_P}=0.076$ at $t<0$, where $T_q^{ex}=8 pN nm$. Then $\sigma_P$ follows the external torque jumping. Subsequently, $\sigma_P$ reaches a new equilibrium defined by $\frac{T_q^{ex}\Omega_0}{c_P}=0.091$ at about $0.1$ ms for 1500 $bp$ and 0.3 ms for 3000 $bp$.      
  
The  torque equilibrium  (Eqs \ref{eq:a3} and \ref{eq:a6}) can be applied to the coexistence state of S and P phases.  The interphase torque equilibrium means the equilibrium at the phase-boundaries, i.e.,  $c_P\sigma_P^{eq}=c_S\sigma_S^{eq}$.     Next, we show that the phase boundaries can be chosen as the slow observables to quantify the plectoneme dynamics.

\subsection*{Two-phase dynamics of DNA supercoiling}   

      We have demonstrated the comparatively fast dynamics of torque to reach equilibrium within the S and P phases, respectively. The fast dynamics can then be averaged out within the two phases.  We show below that the energy gradients at the phase-boundaries drive the transformation between the two phases and thus establish a  two-phase dynamic model for describing the DNA supercoiling.

\begin{figure}[H]
  \centering
    \subfigure[]{
    \label{fig:2:a}
    \includegraphics[width=0.33\textwidth,angle=0]{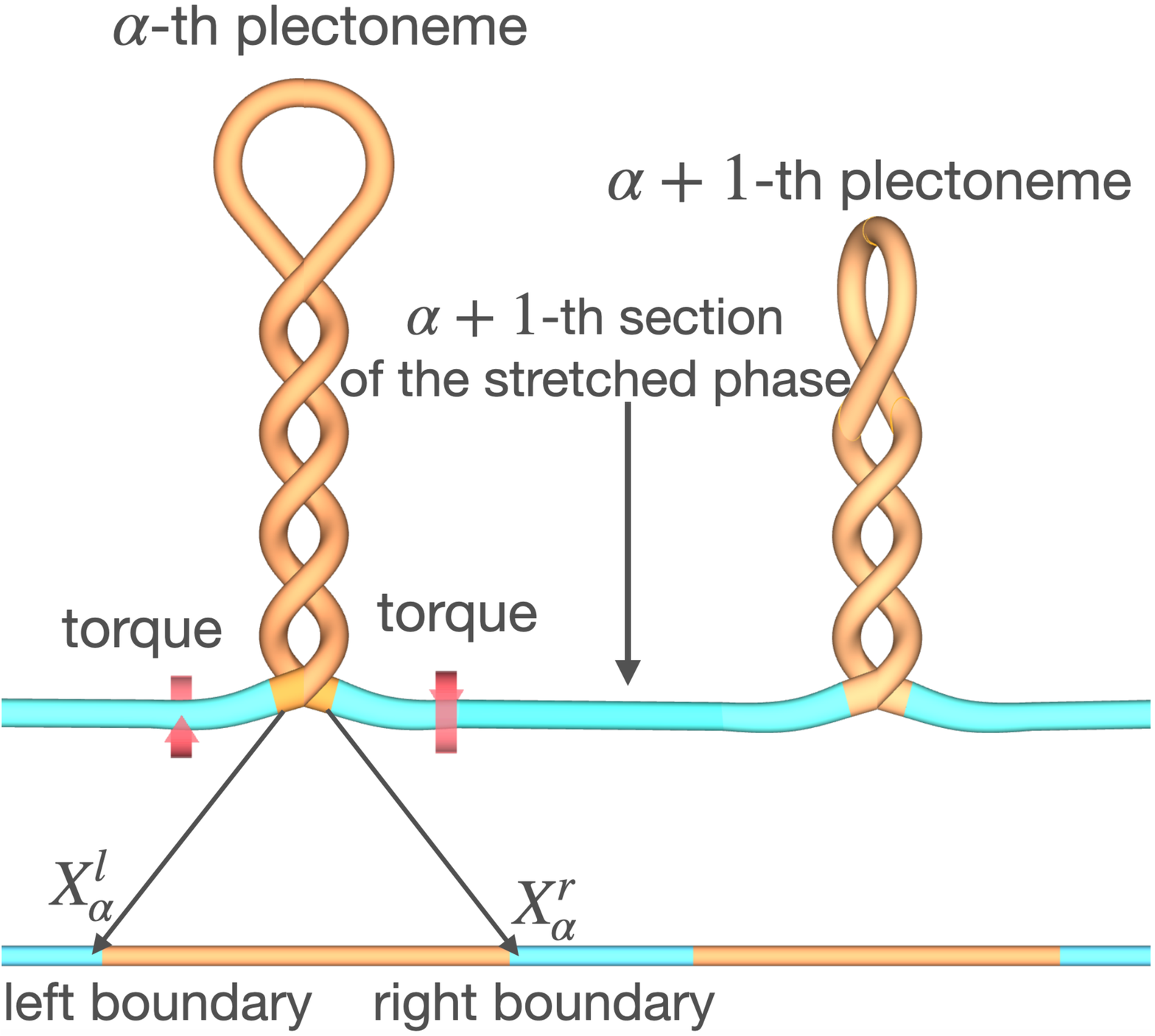}}
    \subfigure[]{
    \label{fig:2:b}
    \includegraphics[width=0.31\textwidth,angle=0]{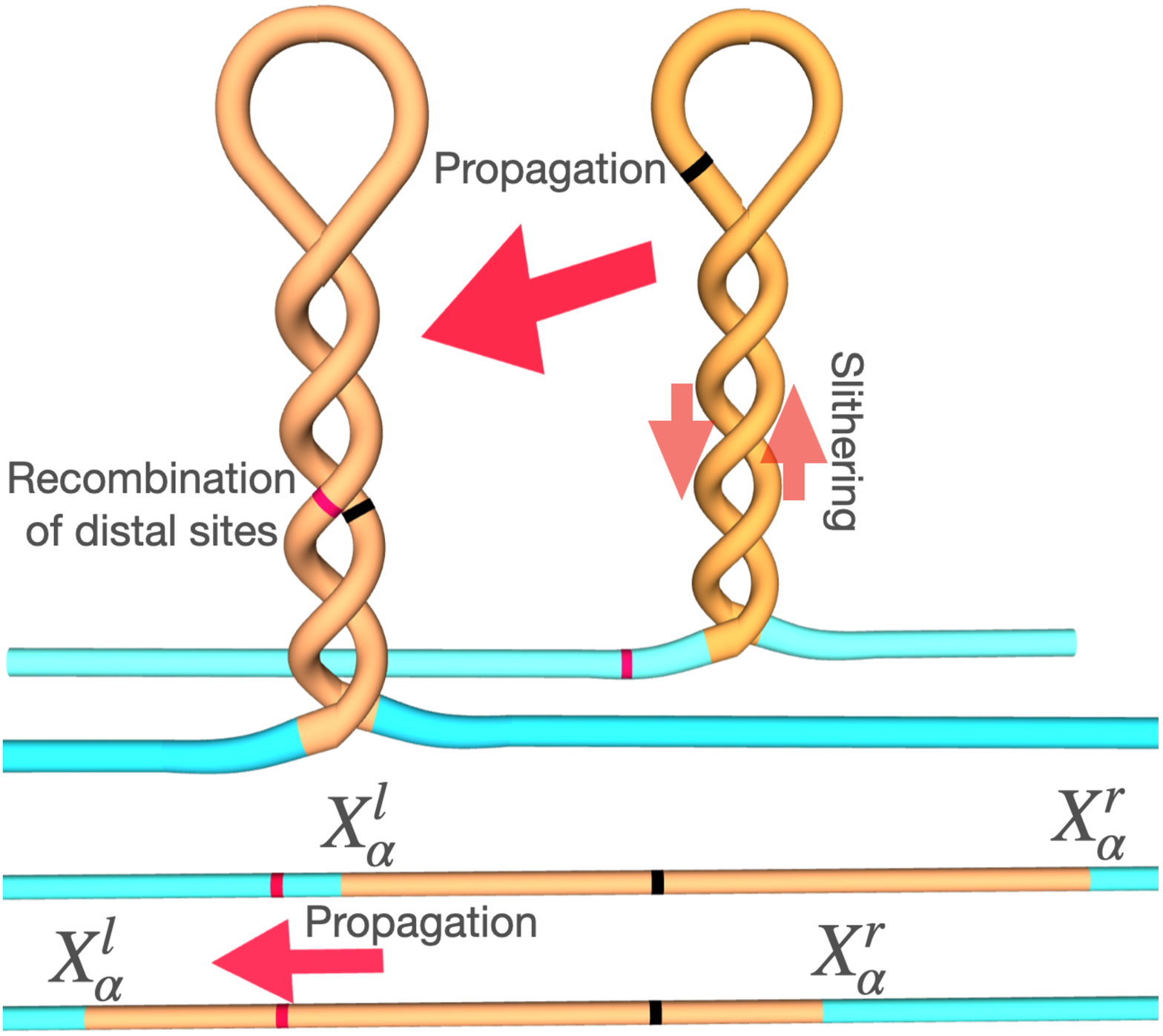}}
        \subfigure[]{
    \label{fig:2:c}
    \includegraphics[width=0.7\textwidth,angle=0]{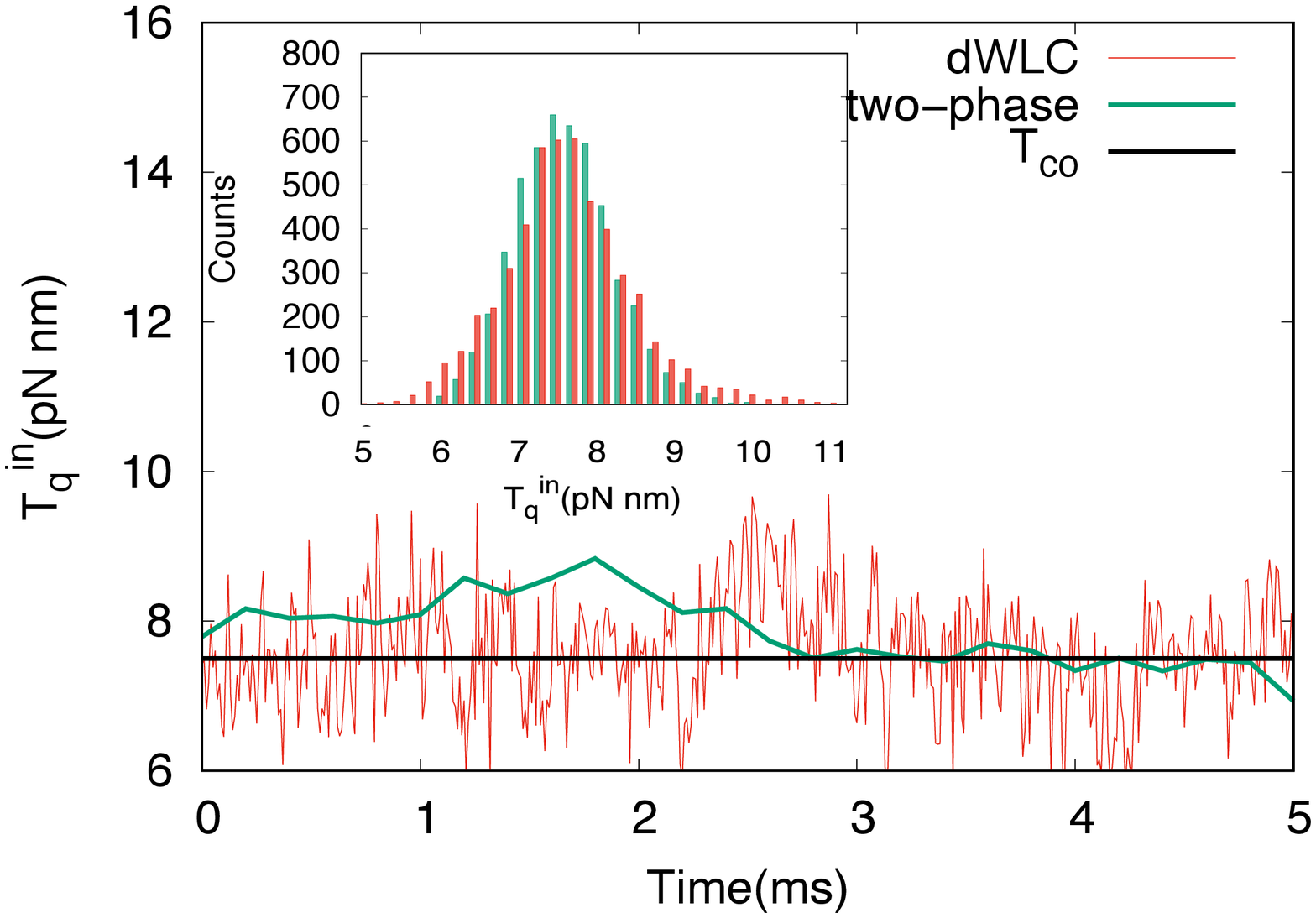}}          
    \caption{ \small Two-phase dynamic model of DNA supercoiling. (a) The schematic of the $\alpha$-th plectoneme among multiple plectonemes, showing  the torque equilibrium between the plectoneme and stretched phases.   By mapping the P and S phases into a 1-dimensional arc-length parameter space, one introduces $u=X_{\alpha}^{l}$, the left boundary of the $\alpha$-th plectoneme, and $u=X_{\alpha}^{r}$, the right boundary of the $\alpha$-th plectoneme.      (b)  The propagation  of the $\alpha$-th plectoneme.  The slithering of the segments of the plectonemic helix facilitates the recombination of distal sites on DNA (marked by red and black marks).  (c)Simulation trajectories based on the WLC model and the two-phase dynamics.  The internal torque of a DNA of 4500 $bp$ at $f=0.5pN$ fluctuates around the coexistence torque, $T_{co}$. The histograms in the inset show the variance of the internal torque obtained from the dWLC simulations (red bars, 50 $ms$ in total) and that from the simulations using the two-phase dynamic model (green bars, 1 $s$ in total).}
  \label{fig:2}
  \end{figure}

     Multiple plectonemes have been observed in the extended DNA supercoil in experiment\cite{van2012Dynamics}.  The dynamics of the plectonemes can be described by the boundaries between the S and P phases.   We therefore label the plectonemes from the fixed end to the stretching end with  $\alpha=1,2, ...$.   Without loss of generality, we only focus on the $\alpha$-th plectoneme as shown in Fig. \ref{fig:2:a}.  ${\mathbf{ X}}\equiv(X_{\alpha}^{l}, X_{\alpha}^{r})$ are the boundaries of plectonemes, where the superscripts $l$ and $r$ denote the left and right boundaries of the $\alpha$-th plectoneme, respectively. Along the axis, $X_{\alpha}^{l}< u \leqslant X_{\alpha}^{r}$ locates the $\alpha$-th plectoneme, while $X_{\alpha-1}^{r}<u<X_{\alpha}^{l}$ specifies the $\alpha$-th section of the S phase. Thus the motion of plectonemes must be accompanied by that of the stretched phase. The formation of plectonemic coils can make  DNA  compact by bringing distant-site together, i.e., $X_{\alpha}^{l}$ and $X_{\alpha}^{r}$ are in physical contact.  Plectoneme propagation corresponds to the movement of $X_{\alpha}^{l}$ and $X_{\alpha}^{r}$ in the same direction. For example, in Fig. \ref{fig:2:b} if the $\alpha$-th plectoneme moves left, the S phase on the left-hand side of  $X_{\alpha}^{l}$ is transformed into the P phase while the P phase on the left-hand side of $X_{\alpha}^{r}$ is transformed into the S phase. This process is accompanied with the internal slithering of parallel segments of the plectonemic helix(Fig. \ref{fig:2:b}). Indeed,
 the velocity of the relative slithering is equal to the velocity of plectoneme motion (see  Methods).  
     
 Such observables always evolve  slowly  comparing to the torque transport, independently of the contour-length (Text S5).     A collective observable $\mathbf{ X}$ is associated with a certain amount of free energy, with its  gradient along $\mathbf{ X}$ as a driving force  to reach  the free energy minima\cite{2011Principles}.  Based on this idea we derive below the free energy and the dynamic equations associated with the slow variable $\mathbf{ X}$.

  \subsubsection*{ Free energy associated with the slow observables }   
 
  As the torque equilibrium is reached, i.e., $\Omega_0\overline{T}_q^{in}=c_P\sigma_P^{eq}$.   The torque equilibrium at the boundaries means  $c_P\sigma_P^{eq}=c_S\sigma_S^{eq}$. Accordingly, every plectenome carries the equal $\sigma_P^{eq}$.  The free energy density of each plectoneme is therefore represented as  $\mathcal{F}_P(\mathbf{ X})=\frac{1}{2}c_P(\sigma_P^{eq})^2$\cite{Klenin1991Computer,1997The}. 

The translational displacement of the $\alpha$-th section of the S phase is subjected to  the viscous drag along it.   At $u<X_{\alpha}^{r}$ and $>X_{\alpha}^{l}$ , the tension imposed on DNA is $f_u$,  and at the stretching end, $f|_{u=L}=f$.  As the torque equilibrium is reached, the local free energy density (or force) is represented by $\mathcal{F}_S(u,\mathbf{ X})=\frac{1}{2}c_S(\sigma_S^{eq})^2+\mathcal{F}_0$\cite{Marko1998DNA,Marko2007Torque}.

The free energy associated with ${\mathbf{ X}}$ can be written as the contributions from the S and P phases 

\begin{equation}
 \Phi_0(\mathbf{ X})=\sum_{\alpha=0}\int_{X_{\alpha}^{r}}^{X_{\alpha+1}^{l}}\mathcal{F}_{S}(u,\mathbf{ X})du+\int_{X_{\alpha+1}^{l}}^{X_{\alpha+1}^{r}}\mathcal{F}_{P}(\mathbf{ X})du. \tag{B1}
 \label{eq:b1}
 \end{equation}

 \subsubsection*{Langevin dynamics} 
The free energy gradient along $\mathbf{ X}$ serves as a driving force, which drives the transformation between S and P phases.  The variable $\mathbf{ X}$ thus follows the Langevin equations   $\gamma_{1}\mathbf{\dot{X}}=-\nabla_{\mathbf{ X}}\Phi_0(\mathbf{ X})+\sqrt{\frac{2k_BT\gamma_1}{dt}}\mathbf{ \dot{W}}$, where $\gamma_1$ denotes the drag coefficent for the growth or shrinkage of plectonemes, and $\dot{W}$ is the white noise.    Specifically, the left/right boundaries of the $\alpha$-plectoneme follow
\begin{equation}
\dot{X}_{\alpha}^{l/r}=-\frac{1}{\gamma_{\alpha}}\nabla_{X_{\alpha}^{l/r}}\Phi_0(\mathbf{ X})+\sqrt{\frac{2k_BT}{\gamma_{\alpha}}}\dot{W},\tag{B2}
\label{eq:b2}
\end{equation}
where $\gamma_{\alpha}$ is drag coefficient for the growth or shrinkage of  the $\alpha$-th plectoneme (see Methods).

 It should be noted that Eq \ref{eq:b2} should be completed with two constraints:  the first one is the topology constraint ($\Delta Lk =\Delta Lk_S+\Delta Lk_P.$), and  the second one is  in-extensibility of DNA model ($L=L_S+L_P$)(see  Methods). 
 
The equilibrium point, i.e.,  $\nabla_{X_{\alpha}^{l/r}}\Phi_0(\mathbf{ X})=0$ leads to the coexistence torque $T_{co}=\frac{1}{\Omega_{0}}(\frac{2c_P(f-\sqrt{\frac{k_BTf}{l_b}})}{1-c_P/c_{S}})^{\frac{1}{2}}$\cite{Marko2007Torque}. Actually, the dynamics of the slow variable $\mathbf{ X}$ accompanies  with a time-dependent torque, $T_q^{in}(t)\equiv \frac{c_P\sigma_P^{eq}(t)}{\Omega_o}$ (Fig. \ref{fig:2:c}).  The internal torque of a DNA at length of 4500 $bp$ under $f=0.5pN$ fluctuates around the coexistence torque, $T_{co}$. The two-phase dynamic model can be regarded as a smoothed version of the dWLC method. Although the fluctuations faster than the torque equilibrium have been averaged out, the mean and the variance of the torque are still consistent with that obtained from the dWLC method(inset in Fig. \ref{fig:2:c}).
 
Supercoil with  a given linking number  is analogous to the classical gas (van der Waals) confined  inside a fixed volume (NVT ensemble)\cite{mattis2008statistical}.    A DNA twisted under a constant torque $T_q^{ex}$, however, is analogous to the classical  gas under  a fixed pressure (isobaric ensemble).     Its free energy densities can be obtained through the Legendre’s transformation $\mathcal{F}_{S/P}(T_q^{ex})=\mathcal{F}_{S/P}-\Omega_0T_q^{ex}\sigma^{eq}$. The free energy densities of the plectonemes and the stretched phase are therefore $\mathcal{F}_{P}=-\frac{1}{2}\frac{(\Omega_{0}T_{q}^{ex})^{2}}{c_P}$ and $\mathcal{F}_{S}=-\frac{1}{2}\frac{(\Omega_{0}T_{q}^{ex})^{2}}{c_S}-f_u+\sqrt{\frac{k_BT f_u}{l_{b}}}$, respectively. Inserting them into Eq \ref{eq:b1} and Eq \ref{eq:b2} we obtained the dynamic model of extended DNA supercoiling under constant torque. 

  Generally speaking, the stretching force, $f$ in Eq \ref{eq:b1} or \ref{eq:b2} is not necessarily constant. The force can be time-dependent  as long as it dose not break the interphase torque equilibrium. For example,  one can impose an instantaneous force.  Similarly, $\Delta Lk$ can be time-dependent as well if it does not break the interphase torque equilibrium.  
   
   \subsection*{Comparing with the WLC model}  
  
  In this section, by employing the dWLC method, we compare and calibrate the two-phase dynamic model  via two examples. One is the supercoiling accumulation at a constantly twisting rate.   The other one is supercoiling accumulation under a constant torque. In the first example,  the system reaches to a two phase-coexistent state at equilibrium. We show that the two-phase dynamic model at trivial computational costs provides consistent results with the dWLC method, which demands a significant amount of computational time.    In the second case, plectonemes finally dominate the supercoiling state  under an  external torque larger than $T_{co}$. This non-equilibrium process helps determine the slithering drag coefficient by calibration.  
  \subsubsection*{Supercoil accumulation and buckling transition at a constant rate}  
  
\begin{figure}[H]
      \centering
    \subfigure[]{
    \label{fig:3:a}
    \includegraphics[width=0.3\textwidth,angle=0]{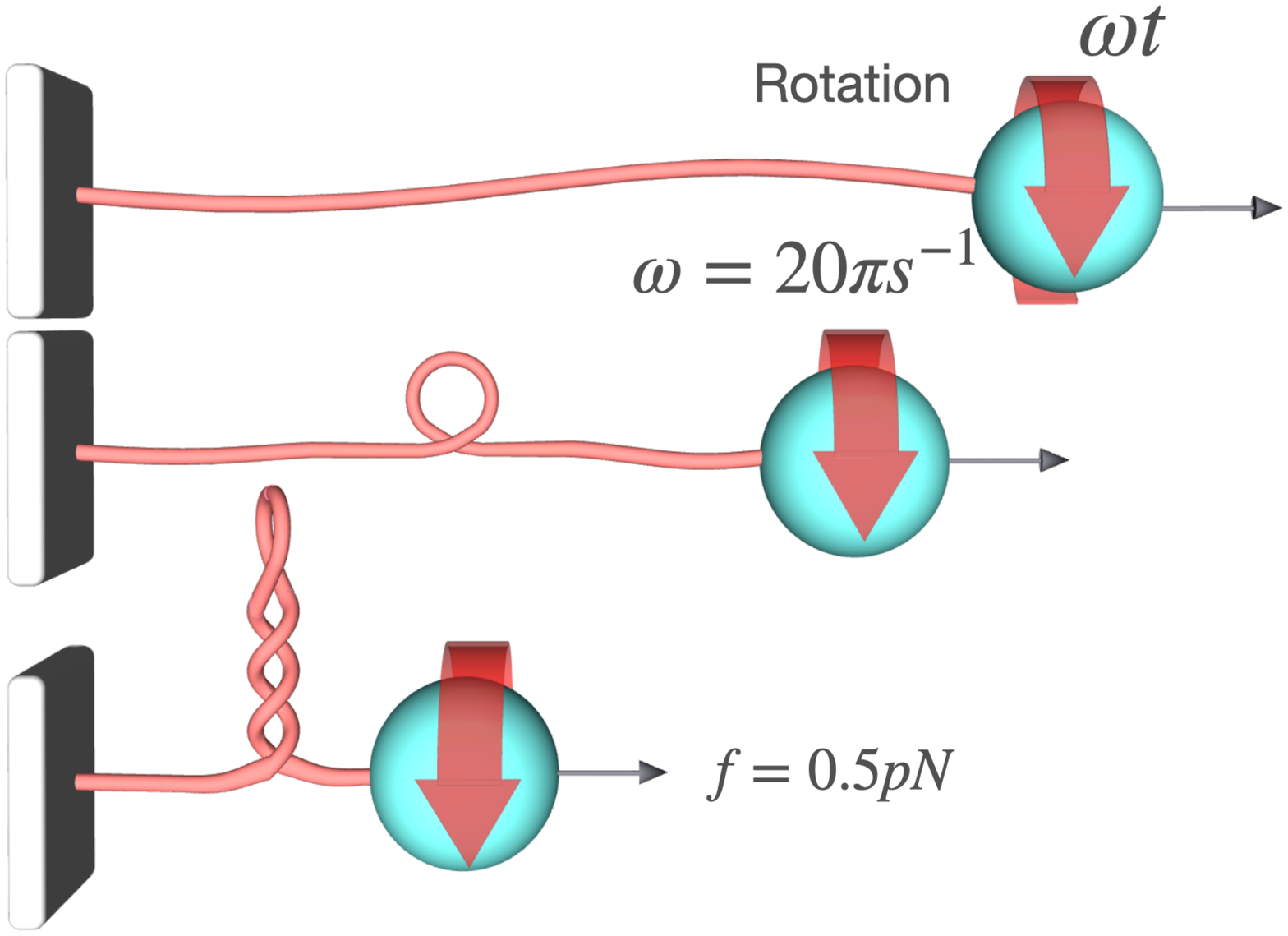}}
        \subfigure[]{
    \label{fig:3:b}
    \includegraphics[width=0.33\textwidth,angle=0]{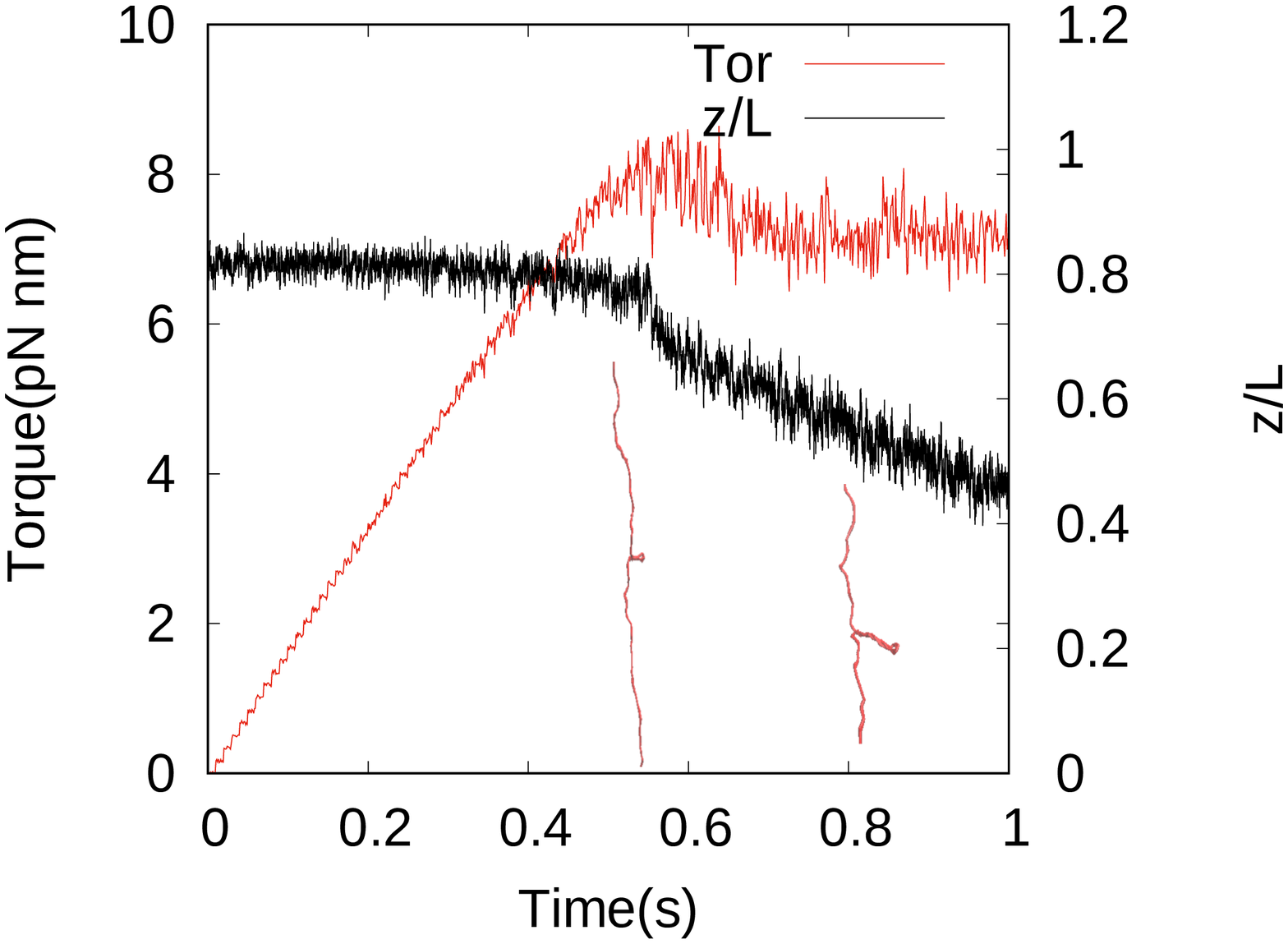}}
    \subfigure[]{
    \label{fig:3:c}
    \includegraphics[width=0.33\textwidth,angle=0]{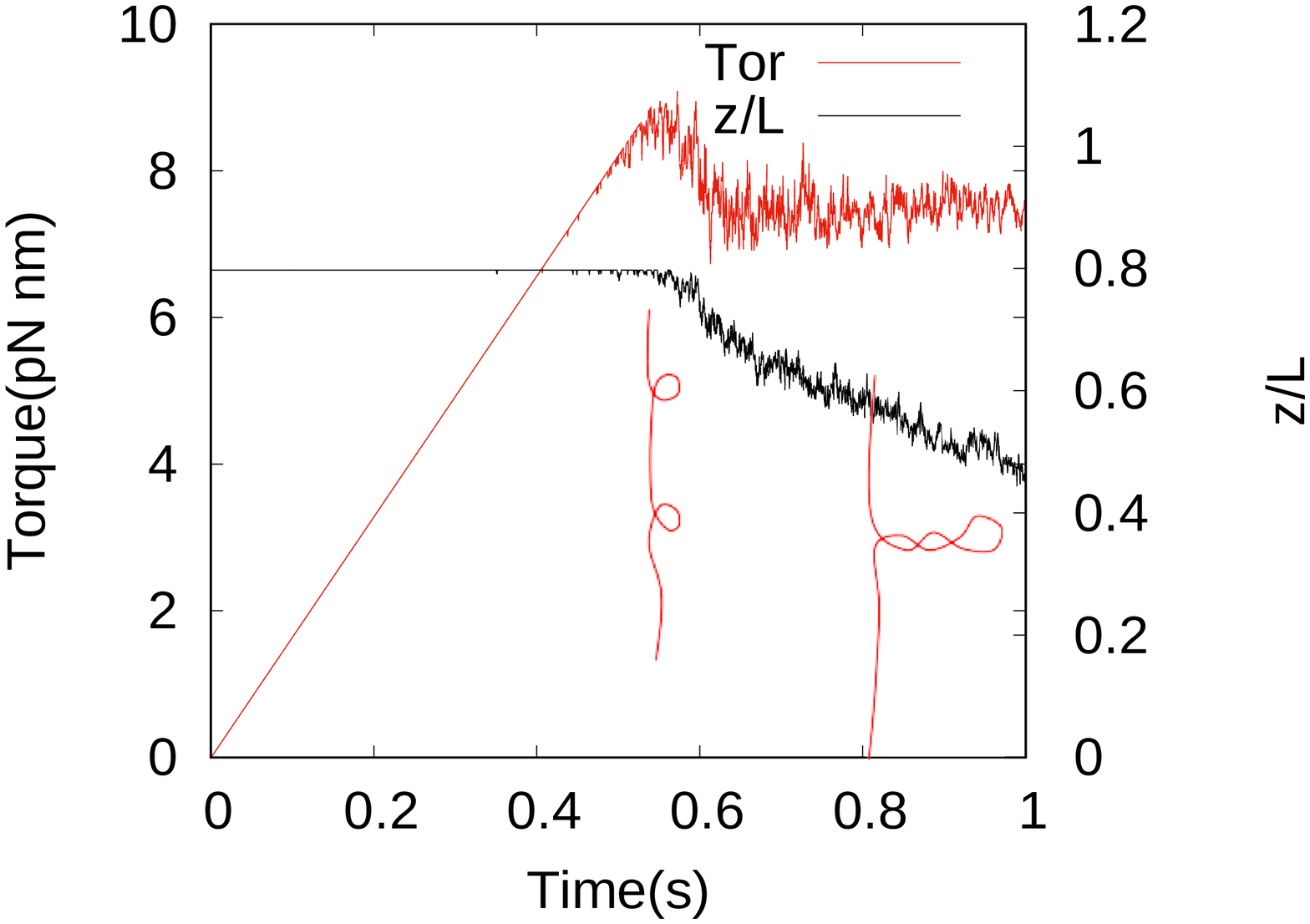}}
        \subfigure[]{
    \label{fig:3:d}
    \includegraphics[width=0.3\textwidth,angle=0]{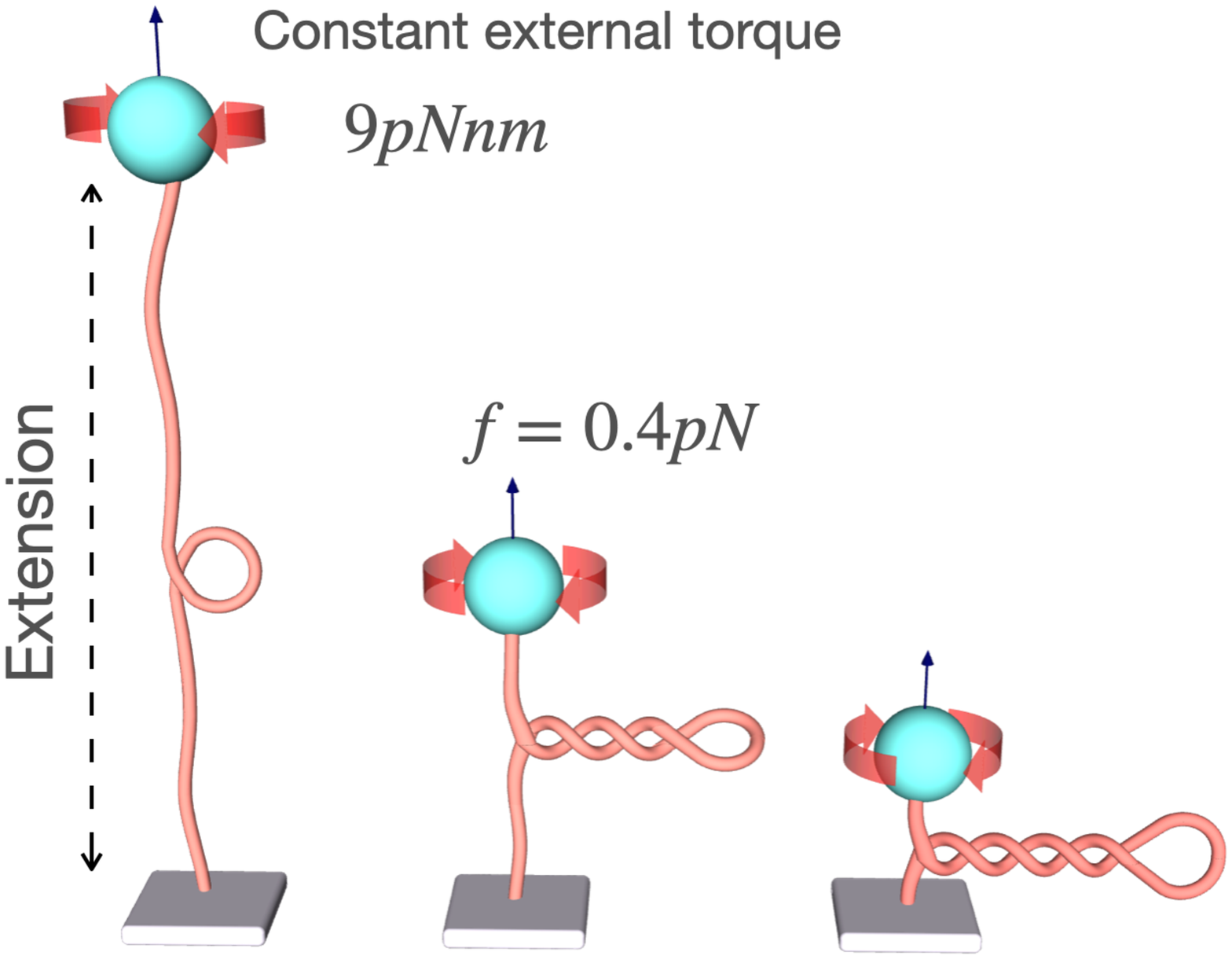}}
    \subfigure[]{
    \label{fig:3:e}
    \includegraphics[width=0.33\textwidth,angle=0]{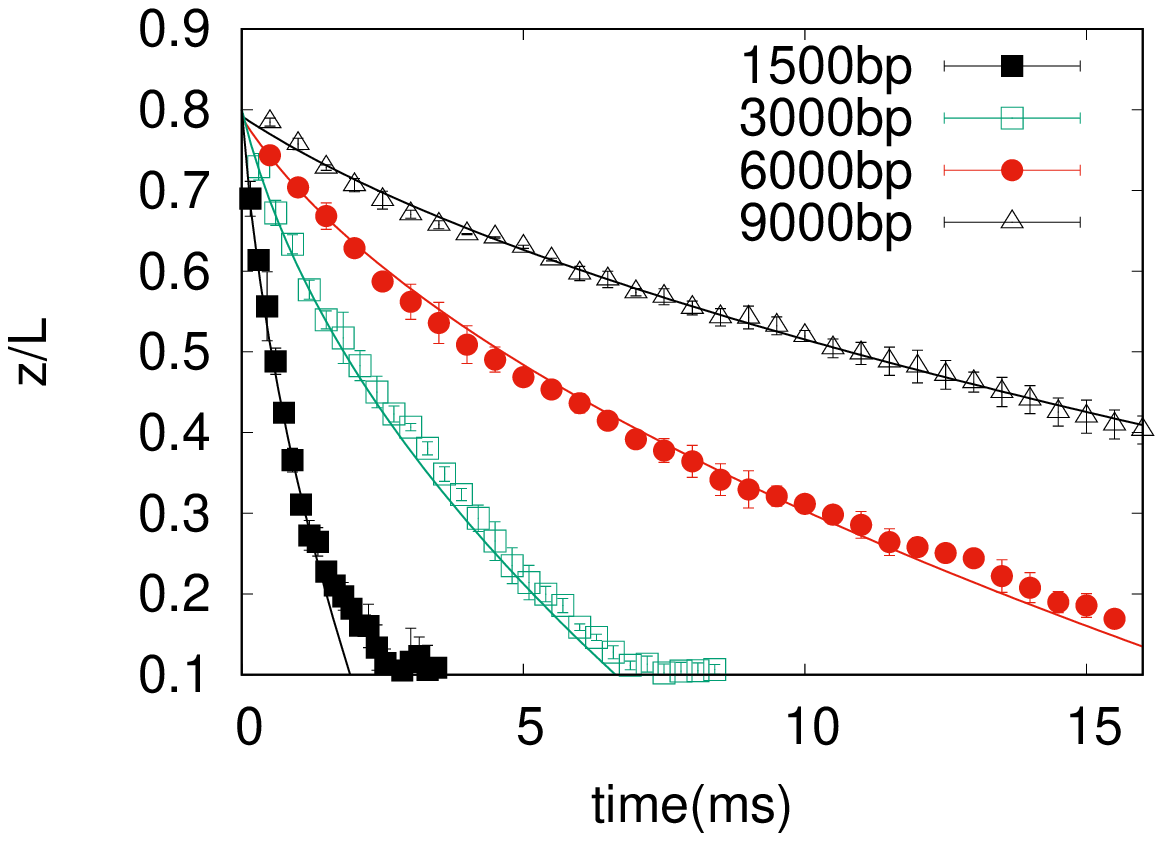}}  
      \subfigure[]{
    \label{fig:3:f}
    \includegraphics[width=0.33\textwidth,angle=0]{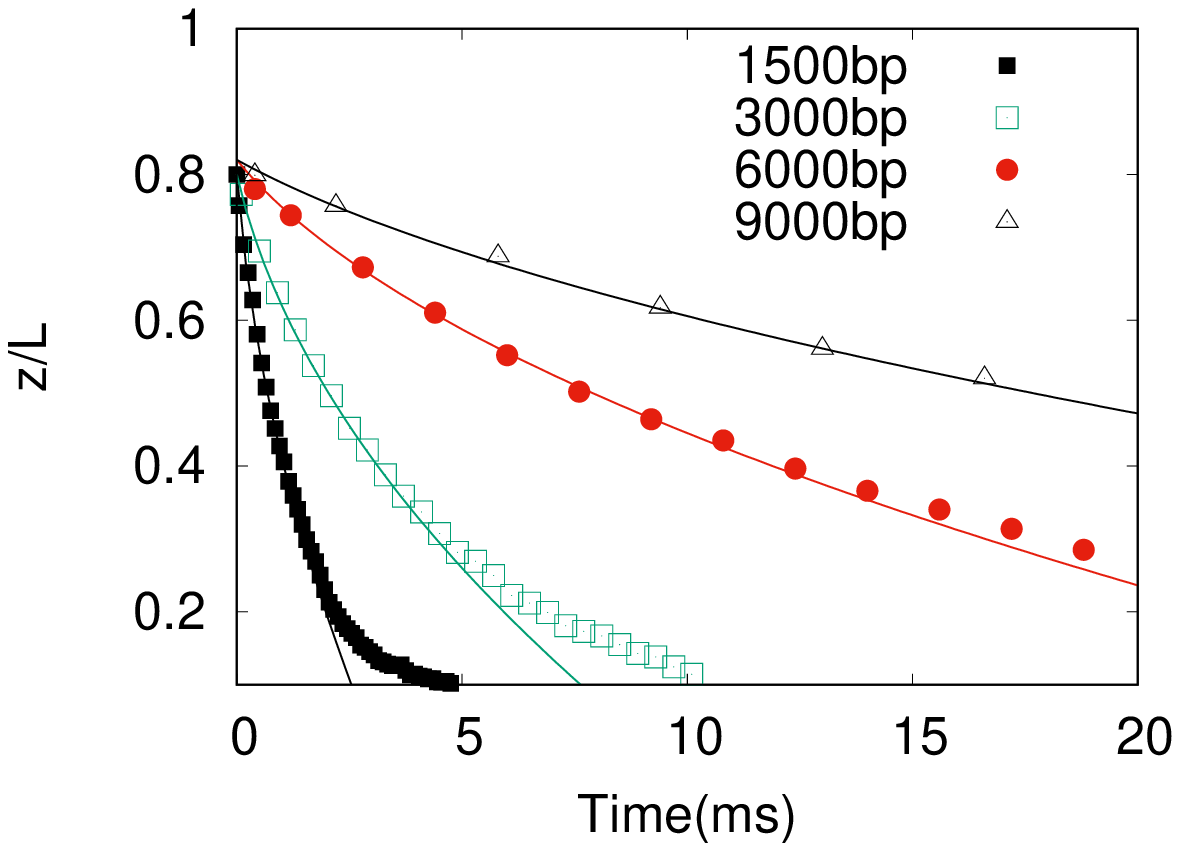}}     
         
    \caption{\small Buckling transition and  measurements of the extensions of the DNA during supercoil accumulation. (a)Injecting supercoil (i.e., increasing linking number) at constant rate to the DNA of 3000 $bp$ within 1 $s$. The rotation rate $\omega$ is 20 $\pi$ per second.    (b)The relative  extension $z/L$ (black curve) and torque (red curve) obtained from the BD simulations based on the dWLC  as the DNA is gradually twisted. (c)The torque and extension curves from (b) are reproduced using the two-phase dynamic model.  (d) Injecting supercoil under a constant external torque, $T_q^{ex}=9 pN nm$. (e)The relative extension  from the BD simulations based on the dWLC with various DNA lengths. (f) The results from (e) are reproduced using the two-phase dynamic model.   The curves in (e) and (f) are the fitting functions $\sqrt{1+\lambda t}$ (see  Text S7). }
  \label{fig:3}  
  \end{figure}

  An extended  DNA undergoes linking number accumulation when one end of the DNA is rotated while the other end is kept fixed,  as shown in Fig.\ref{fig:3:a}.  The stretching force $f=0.5 pN$, the rotation angle  $\Omega=\omega t$ and the angular velocity  $\omega=20\pi/s$, equivalent to $\Delta Lk=10$ per second (i.e., comparable to polymerase enzyme unwinding and synthesizing at ~100 nucleotide or nt per second)\cite{thomen2005unravelling,Jing2018How}.  At first,  the extended DNA is twisted, and then it undergoes a buckling transition (discontinuities on torque and extension curves) followed by plectoneme formation\cite{Forth2008Abrupt}.   
  
    We  performed BD simulations of such a supercoil accumulation based on the dWLC method. In Fig. \ref{fig:3:b}, the  relative  extension $z/L$  almost remains unchanged before 0.6 s while the torque gradually builds up, suggesting that the extended DNA is twisted.  The buckling  at about 0.6 s marks the starting-point of the phase-coexistence. The buckling transition indicates a competition between  pre-plectonemic loops and stretched  phase\cite{Marko2012Competition,Mielke2004Transcription}.   After 0.6 s, the torque drops and converges  as torsional stress on DNA is relieved,  indicating the twist is transformed into writhe.   These discontinuities have previously  been observed in single-molecule experiment\cite{Forth2008Abrupt}.   The coexistence torque $T_q^{co}$ remains unchanged when more linking number is injected. Note that multiple plectonemes can co-exist in the process. The representative snapshots are also shown in Fig. \ref{fig:3:b}.   The unchanged torque curves or the linearly decreased extension after buckling transition suggest that the process remains quasi-static.
    
 It should be noted here that we obtained the curves in Fig. \ref{fig:3:b} from four BD simulation trajectories using the dWLC method, and each of the trajectories requires several weeks of CPU time. The calculation of the electrostatics interaction between the charged points on segments of the dWLC model indeed dominates the computational cost (see Text S6). According to our analyses, the computational cost of the dWLC in the plectonemic phase is still significant and grows  linearly with the contour-length. For example, in Fig. \ref{fig:1}, simulating a stretched DNA of 3000 $bp$ in 10 $ms$  costs 3 hours while simulating a pure plectoneme of the same contour length costs one day(i.e., the CPU time).  The two-phase dynamic model, however, enables us to  reduce the computational cost to trivial by dealing only with the essential variables with slow dynamics.   In Fig. \ref{fig:3:c}, the torque and extension curves generated from the two-phase dynamic model (using tens of seconds of computation time) well reproduce those in Fig.3(b) from the dWLC method and also consistent with previous experimental and computational studies\cite{Forth2008Abrupt,ott2020dynamics}.  
  
  \subsubsection*{Supercoil accumulation under a constant torque}

 We also performed simulations of supercoiling generation under a constant torque. Fig.\ref{fig:3:d} illustrates an extended supercoil under a constant torque, $T_q^{ex}=9 pN nm$ (comparable to 5$\sim$11 $pN nm$ torque exerted by polymerases\cite{Ma2013Transcription}), and a constant stretching force $f=0.4 pN$. Since the external torque is larger than $T_{co}$ (i.e., $6.5 pN nm$ that maintains the coexistence state under 0.4 $pN$), plectonemes are created and the ultimate equilibrium state is dominated by plectonemes. Consequently,   plectonemic coils are building up until all stretched DNA is interwound to plectonemical coils.
   
Using the dWLC method, we can capture the dynamics of the plectoneme growth.    Fig. \ref{fig:3:e} shows the  measurements of   the relative extensions $z/L$ of DNAs at 1500 $bp$ (black solid squares), 3000 $bp$, 6000 $bp$ and 9000 $bp$. Indeed, the shrinking extensions indicate the growth of plectonemes.  		The fitting curves are  generated using $\frac{z}{L}-\frac{z_0}{L}\propto1-\sqrt{1+\lambda t}$, where $\lambda$ is a fitting parameter.   In the two-phase dynamic model, all but the slithering  drag coefficients are given based on the hydrodynamics (see Methods).  The slithering drag $\mu_{slt}$ is to be determined in the two-phase model by calibrating results with the dWLC method.   We  performed the simulations using the two-phase dynamic model shown in Fig. \ref{fig:3:f}.  The slithering drag $\mu_{slt}$ was then determined via fitting parameter $\lambda$ Text S7).   Consequently, we can then reproduce the plectoneme dynamics.

  \subsection*{Reproducing experimentally measured plectoneme dynamics}  
 
 Previous experimental studies show that the plectonemes on the extended supercoiled DNA can diffuse, vanish and re-emerge,  and their dynamics depend on the stretching force and ionic strength\cite{van2012Dynamics}.  
    \begin{figure}[H]
      \centering
    \subfigure[]{
    \label{fig:4:a}
    \includegraphics[width=0.3\textwidth,angle=0]{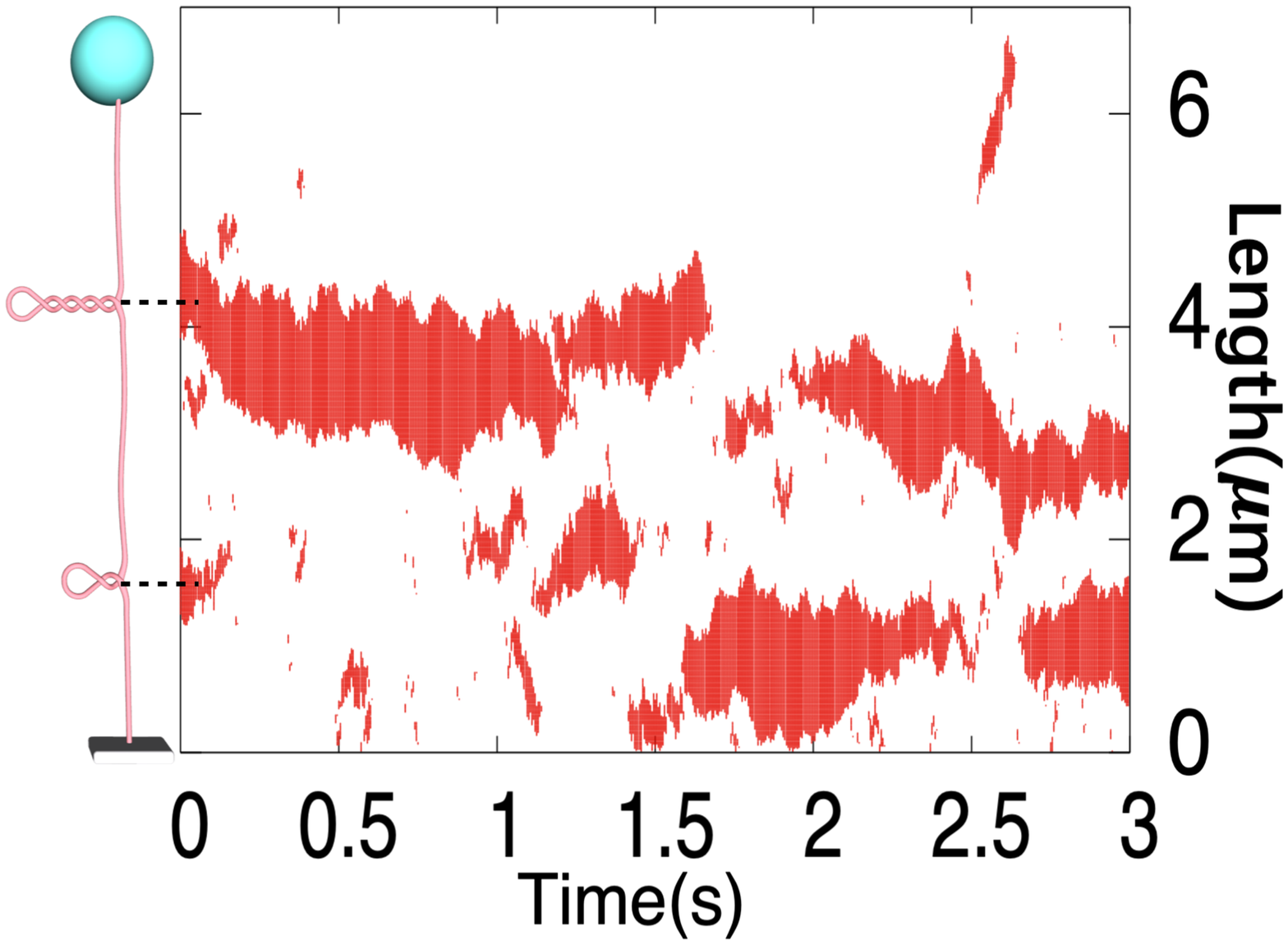}}
        \subfigure[]{
    \label{fig:4:b}
    \includegraphics[width=0.3\textwidth,angle=0]{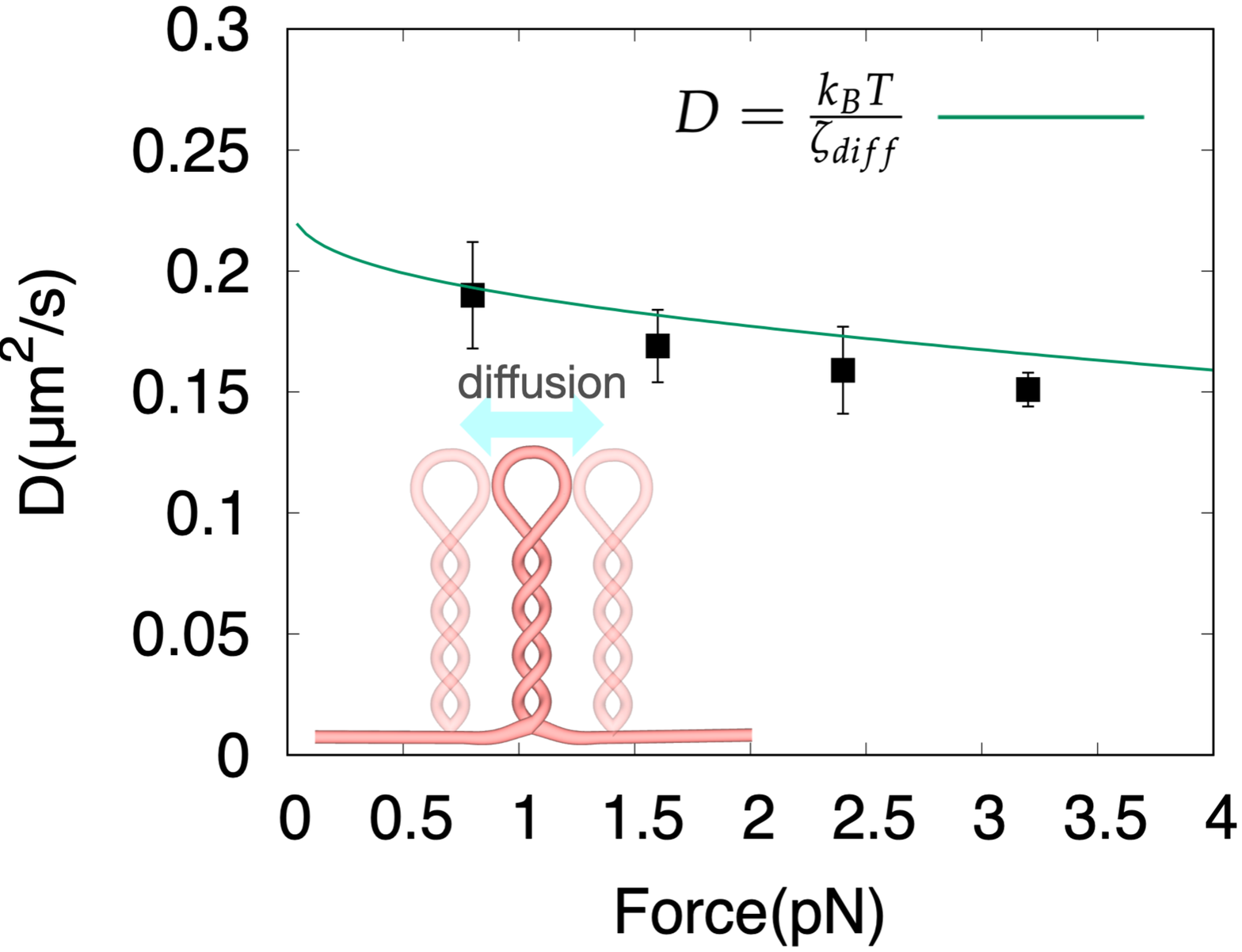}}
       \subfigure[]{
    \label{fig:4:c}
    \includegraphics[width=0.25\textwidth,angle=0]{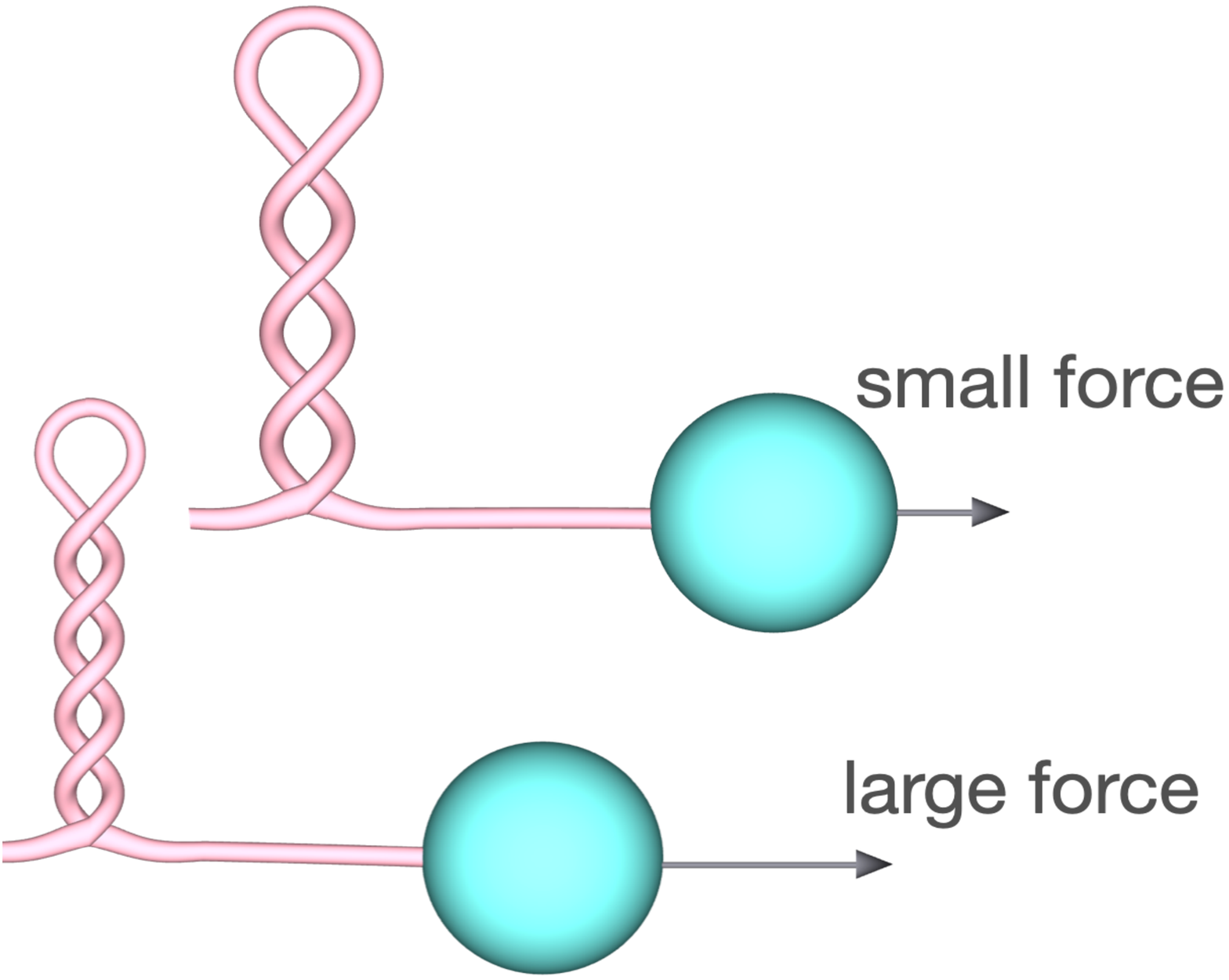}}
        \subfigure[]{
    \label{fig:4:d}
    \includegraphics[width=0.25\textwidth,angle=0]{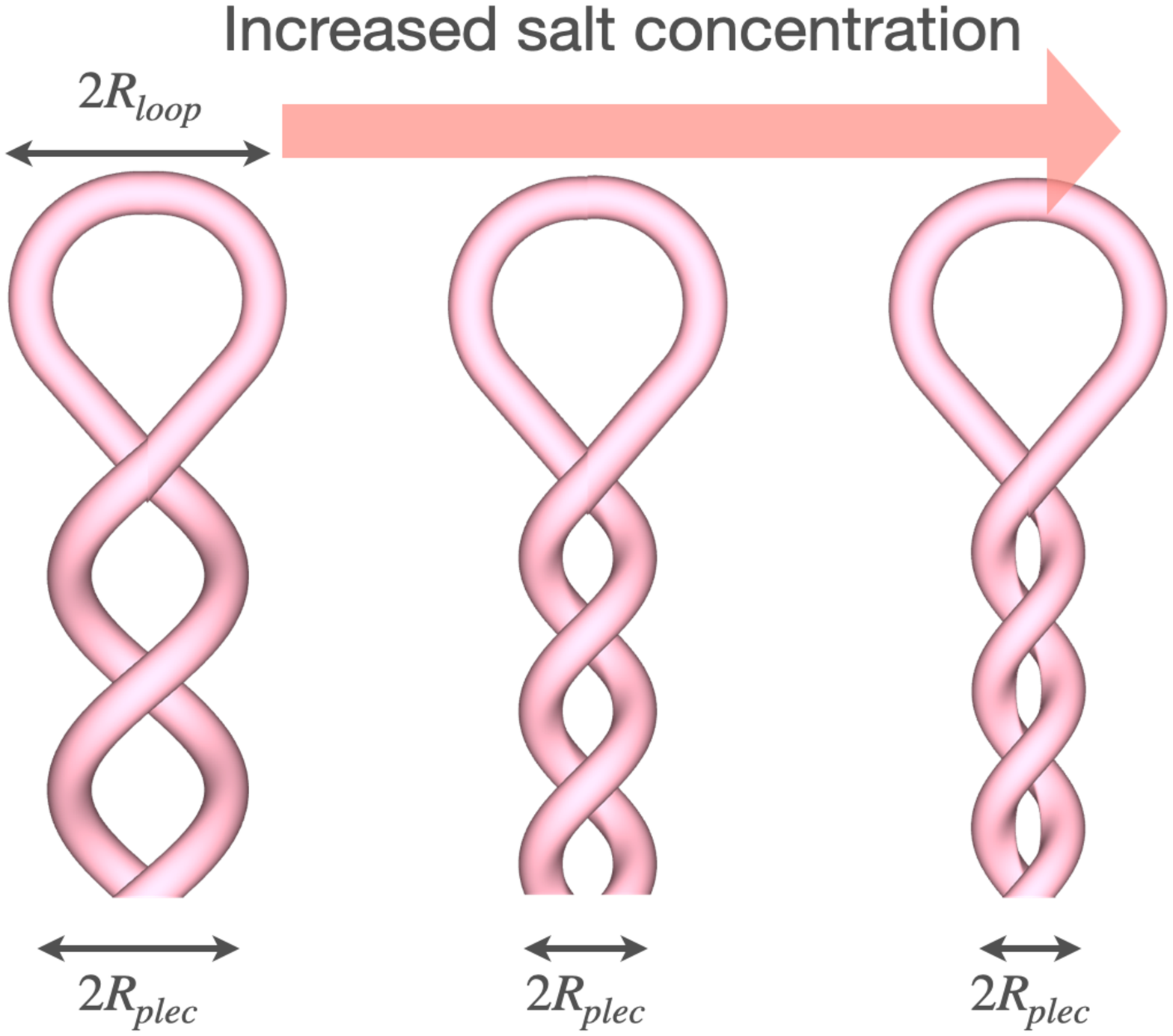}}
        \subfigure[]{
    \label{fig:4:e}
    \includegraphics[width=0.32\textwidth,angle=0]{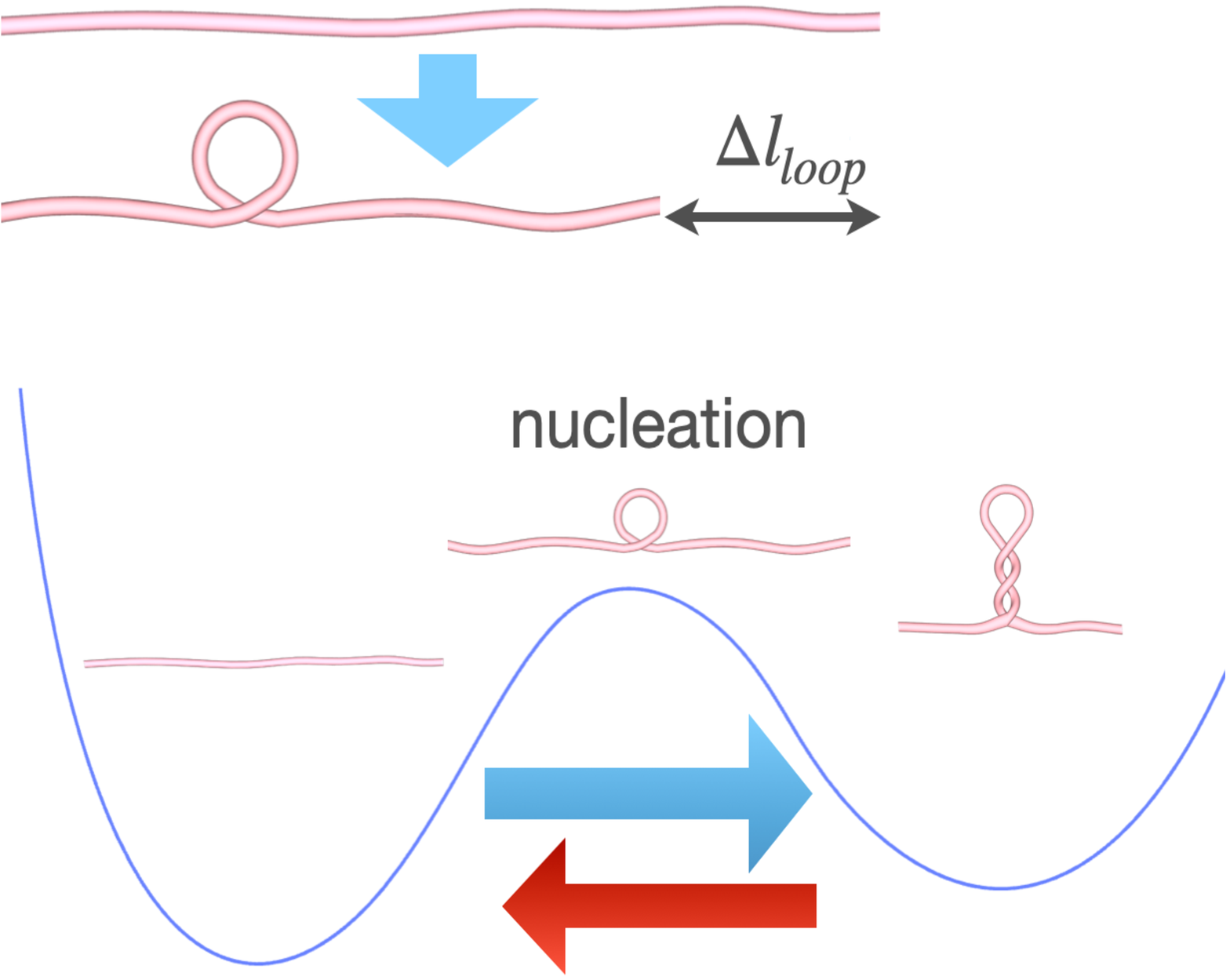}}    
         \subfigure[]{
    \label{fig:4:f}
    \includegraphics[width=0.3\textwidth,angle=0]{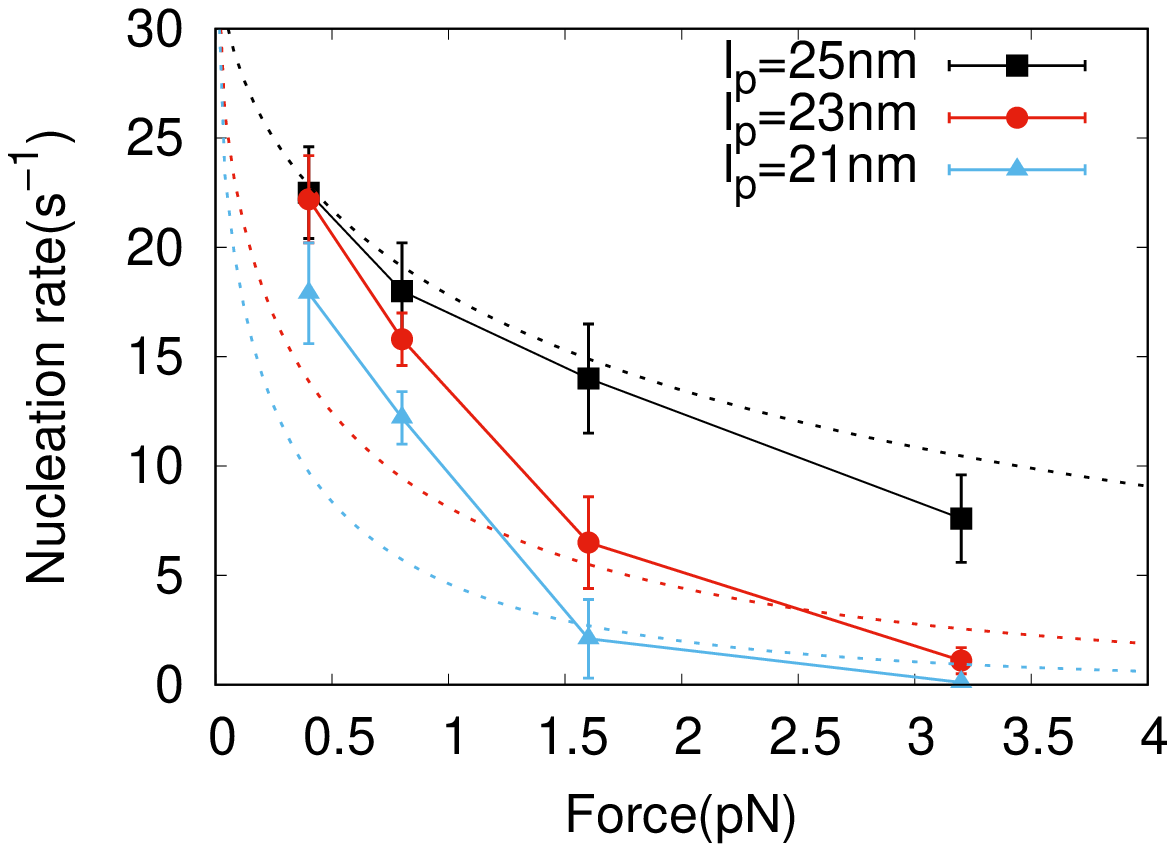}}
        \subfigure[]{
    \label{fig:4:g}
    \includegraphics[width=0.3\textwidth,angle=0]{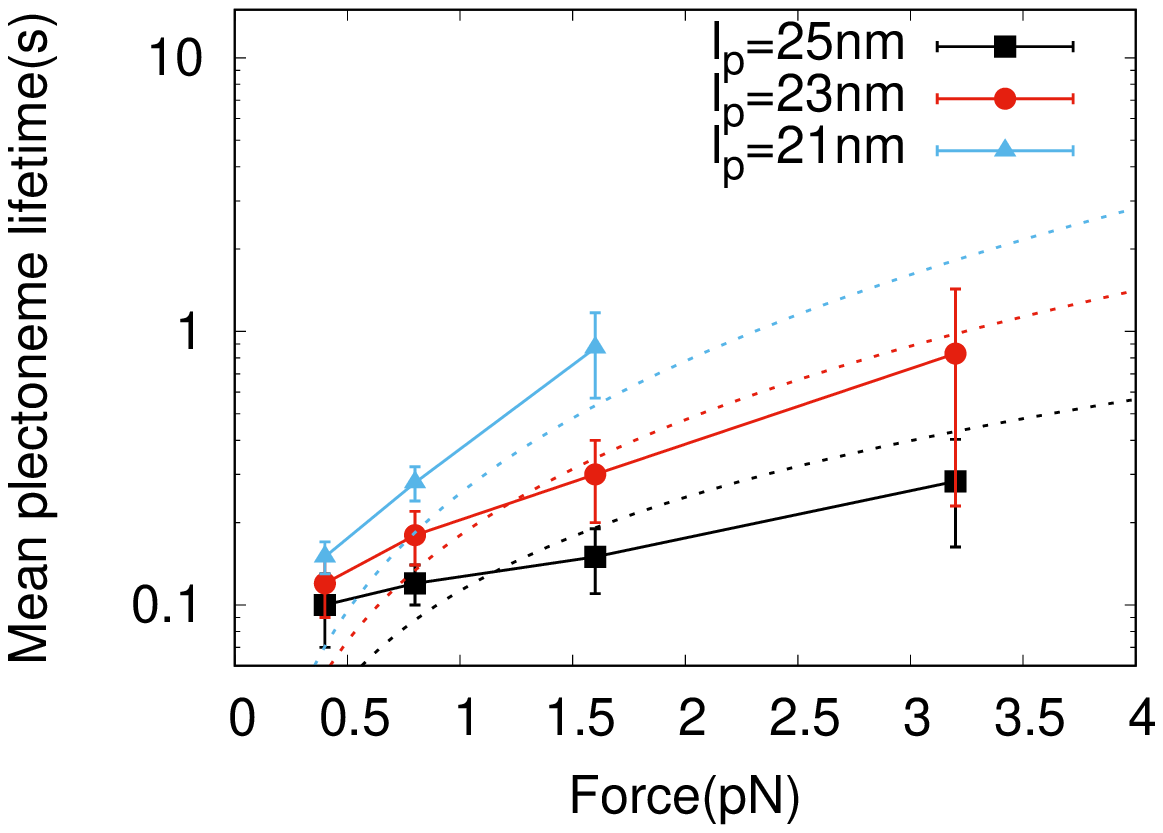}}
    \subfigure[]{
    \label{fig:4:h}
    \includegraphics[width=0.3\textwidth,angle=0]{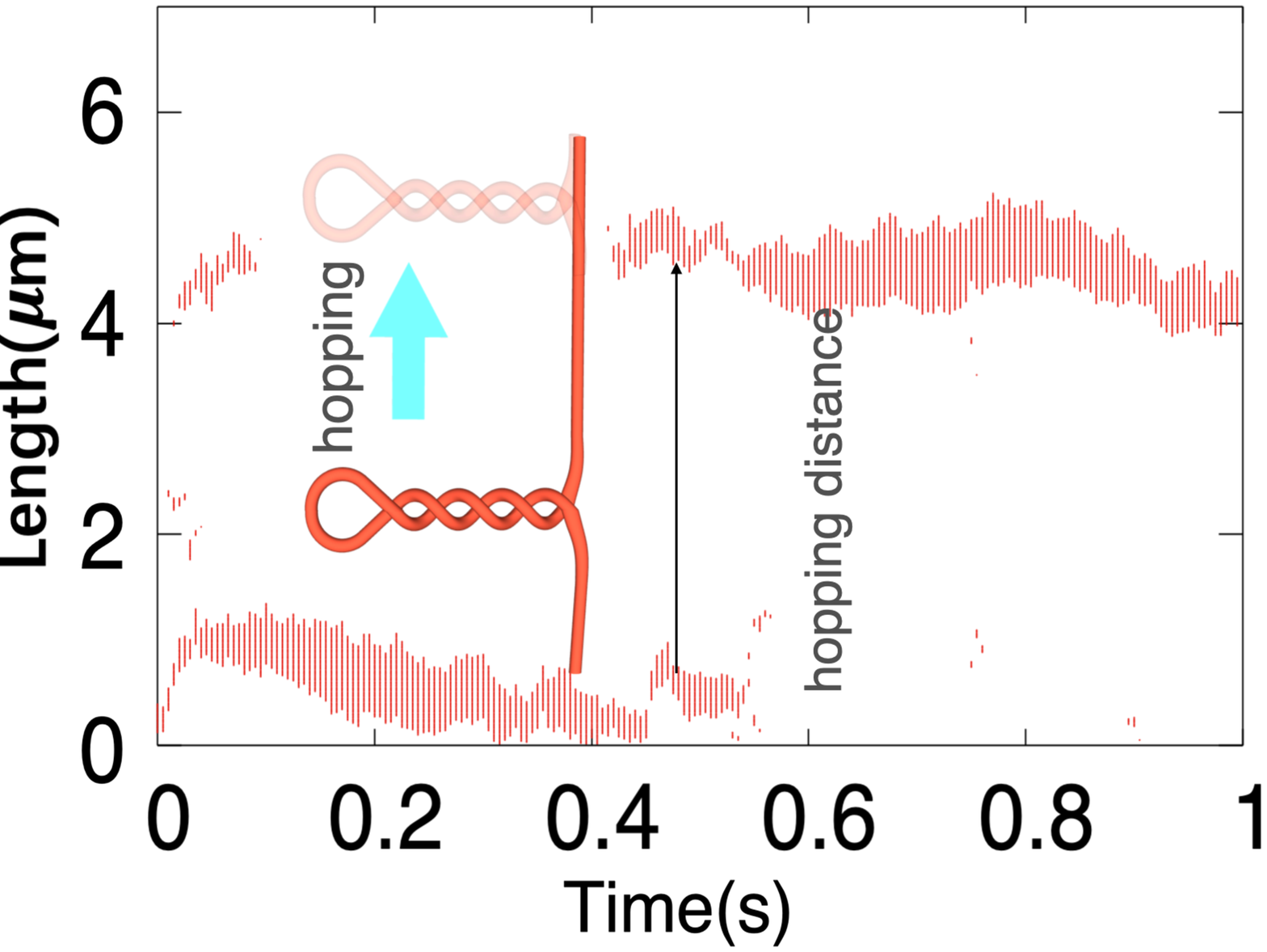}}  
      \subfigure[]{
    \label{fig:4:i}
    \includegraphics[width=0.3\textwidth,angle=0]{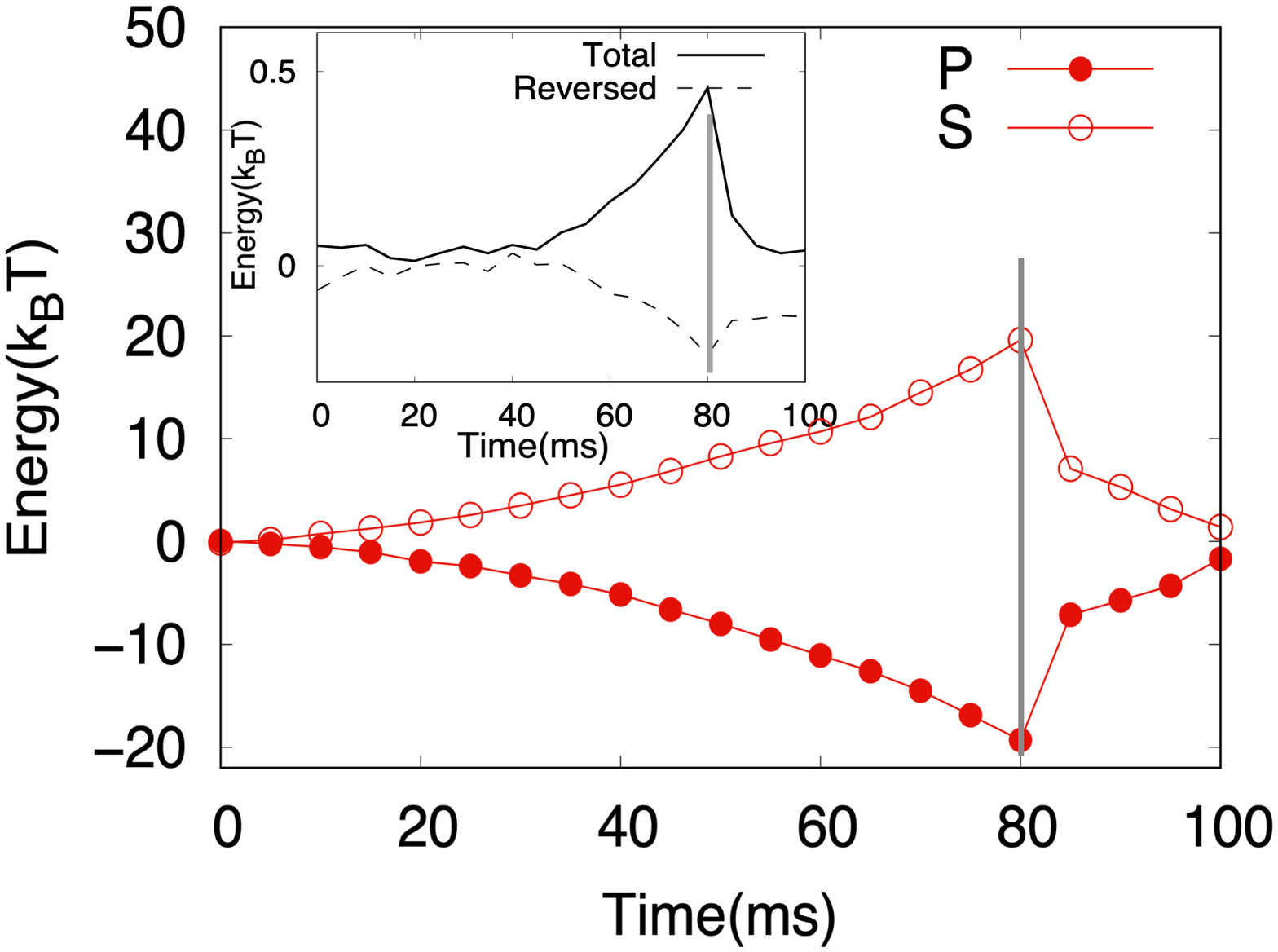}}   
    
      \caption{\small  Obtaining plectoneme dynamics using the two-phase dynamic model. (a) The kymograph of the plectonemes of a 21-kbp (7$\mu m$) under an constant stretching $f=0.8pN$ and with a constant linking number $\Delta Lk = 70$. (b)The diffusion coefficients of a single plectoneme under different forces obtained from the simulation of the two-phase model (the dots). The green curve is directly from the Einstein relation.  (c) Large stretching force $f$ makes more compact plectonemic helices (smaller radius of the plecotnemic helix). (d) Increased salt concentration makes more compact plecotnemic helices. (e) Buckling transition. A portion of stretched DNA is bent to form a loop during nucleation, which is associated  with an energy barrier. The reversed process (red arrow, a plectoneme degenerates to a loop) to the nucleation (blue arrow) is also illustrated.     (f) Nucleation rate of the plectonemes as a function of force obtained from the  two-phase model simulations. The black, red and blue curve and error bars are the results for ($l_p=25 nm$), ($l_p=23 nm$) and  }   
     \label{fig:4}
\end{figure}     
  {\small($l_p=21 nm$) corresponding to the low, regular and high salt concentrations, respectively.   (d) The mean lifetime of the individual plectoneme as a function of force and salt concentration.  Note that only a single plectoneme is captured at $l_p=21 nm$ and $f=3.2 pN$  in simulations (nucleation and annihilation, however, need multiple plectonemes).  (h) A typical hopping event of  a displacement of plectoneme for nearly 4 $\mu m$ in a time interval about $0.1 s$, corresponding to 12 kbp.  (i) Energetic conversion during hopping ( nucleation is, e.g, at 80 $ms$). The total energy changes of hopping (black line) and the reversed (complementary to hopping) process (dashed line) are also shown in the inset.}  
  
 In particular, the experiments also captured  plectoneme hopping, in which a plectoneme  vanishes at one position and a new one immediately emerges  by nucleation at another remote position.    
By employing the  two-phase dynamic model,  we reproduced the plectoneme dynamics, including plectoneme diffusion, nucleation and hopping.  Notably,  we captured the plectoneme hoppings with significant energetic changes, which facilitate such rapid processes.

   \subsubsection*{Reproducing plectoneme diffusion}

  The inter-phase boundaries in the two-phase dynamic model can be used to directly monitor the plectoneme dynamics. Fig. \ref{fig:4:a} shows  a kymograph, i.e., the spatial vs time evolution process  of the plectonemes of a 21-kbp (7$\mu m$) DNA  (extended under  a force $f=0.8pN$ and  with $\Delta Lk = 70$), illustrating the coexistence of multiple plectonemes, and the plectoneme nucleating  and vanishing.

  The simulations of a plectoneme with $1/4$ of the DNA contour length under different forces were performed. To estimate the diffusion constant, we measured the mean square displacement (MSD) $\overline{\Delta x^{2}(t)}$  of the plectoneme.   As shown in Fig. \ref{fig:4:b}, we obtained the diffusion constants via $D\equiv \frac{\overline{\Delta x^{2}(t)}}{2t}$ under different stretching forces.  For consistency check, we also compared the results with the Einstein relation, i.e., $D=\frac{k_BT}{\zeta_{diff}}$ where $\zeta_{diff}$ as a sum of  the viscous drags for  the transverse displacement of the plectoneme (see  Methods).  The diffusion constant of the plectoneme is about $10^{-1}\mu m^2/s$.

  \subsubsection*{Reproducing  nucleation and  Measuring plectoneme life time}
  
  To compare the dynamics of the plectonemes at different stretching forces and ionic strengths,  simulations of the P phase with $1/4$ of the DNA contour length were performed  in the conditions where the stretching forces were $0.4$, $0.8$, $1.6$ and $3.2$ $pN$.      Large stretching force $f$ results in more compact plectonemes, i.e., reduces the radius of the plectonemic helix(Fig \ref{fig:4:c}).  Indeed, in the coexistence state of DNA supercoiling, the equilibrium excess linking number  density of the P phase is $\sigma_P^{eq}=\frac{1}{c_P}(\frac{-2c_P(f-\sqrt{\frac{k_BTf}{l_b}})}{1-c_P/c_{S}})^{\frac{1}{2}}\sim\sqrt{f/l_p}$\cite{Marko2007Torque}, which suggests that $\sigma_{P}^{eq}$ increases with $f$.

  At high salt concentration,  the screening of the Coulomb repulsion potential reduces the effective repulsive DNA diameter, leading to the shorter torsional persistence length of the plectoneme $l_p$ and smaller radius of the plectonemic helix\cite{1997The}. Meanwhile, based on the coexistence state relation $\sigma_{P}^{eq}\propto \sqrt{f/l_{p}}$, lowering $l_{p}$ increases the 
  $\sigma_{P}$, i.e., making the plectonemes more compact (Fig. \ref{fig:4:d}). Accordingly, we adjusted the torsional persistence length of plectoneme $l_{p}$ to represent the impacts from the ionic strength\cite{1997The}, specifically,  $l_{p}=25$, $23$ and $21$ $nm$ to represent the low, regular and high salt concentrations, respectively.  
  
 According to Kramers’ reaction-rate theory, the nucleation rate  is $k_{nuc}\propto exp(-\beta \epsilon_{nuc})$ \cite{Daniels2011Nucleation}, where the energy barrier $\epsilon_{nuc}\propto\sqrt{f}$,  which incorporates the bending energy of the loop and the loss of the elastic energy stored originally in the stretched DNA(Fig. \ref{fig:4:e})\cite{Marko2012Competition}.  Therefore, the large forces can lower the nucleation rate.  Again, according to the relation $\sigma_{P}^{eq}\propto \sqrt{f/l_{p}}$, large force and small $l_{p}$ (high salt concentration) have a similar effect on the nucleation rate. Indeed, the nucleation rates obtained from simulations  (Fig. \ref{fig:4:f}) suggest that large stretching forces $f$ and the high salt concentrations (small $l_p$) can suppress the nucleation events.    We derived the nucleation relation (Text S8)
\begin{equation}
k_{nuc}\propto exp(-2\sqrt{f}C(l_p^{\star}-l_p)),\tag{D1}
\label{eq:d1}
\end{equation}
  where $2C$ and $l_p^{\star}$ are fitting parameters (dashed curves in Fig. \ref{fig:4:f}).   As $l_p\rightarrow l_p^{\star}$ (salt concentration decreases), the nucleation rate increases.  That is because low salt concentration increases the radius of the plectonemic helices i.e.,$R_{plect}\rightarrow R_{loop}$ (bending energy of DNA in plectonemes increases), then lower the energy barrier from plectonemic helix to loop (reversed buckling transition, red arrow in Fig. \ref{fig:4:e}).  By fitting, we obtained $l_p^{\star}=27 nm$.

  The results above indicate that the plectonemes frequently vanish and re-emerge (via nucleation), which lead to the hopping. The surviving time of the individual plectoneme from nucleation to vanishing  is defined as plectoneme life-time. Fig. \ref{fig:4:g} shows that the mean life-time  $\tau_{plec}$ depends on stretching force and salt concentration. The large $f$ or small $l_{p}$ increases the plectoneme life-time.   As discussed above, large $f$ or small $l_{p}$ lowers the plectonemes nucleation and also leads to more compact plectonemes. The resulting  compactness stabilizes existing plectonemes, which means long mean life-time of the individual plectonemes.  By obtaining the balance between the  plectoneme nucleation and vanishing,   we can derive a mean life time of the individual plectoneme (see Text S8)
  
    \begin{equation}
 \tau_{plec}\propto (2l_{t}-l_p)f exp(C\sqrt{f}(l_p^{\star}-l_p)).\tag{D2}
 \label{eq:d2}
 \end{equation}
  where $C$ and $l_p^{\star}$ are the same as those in Eq \ref{eq:d1}.

  \subsubsection*{Plectoneme hopping and the associated energetic changes}

 An example of hopping $over 10 kbp$ within about $0.1s$ is shown in Fig.\ref{fig:4:h}.  Plectoneme hopping is a special case of the DNA conformational rearrangement\cite{van2012Dynamics}. As addressed above, plectoneme diffusion happens comparatively  slowly, only hundreds of bps per second  ($\sim\sqrt{2Dt}$).  Hopping , however, is a rapid process, potentially, facilitating fast site-specific recombination and enhancer activated expression\cite{Sheinin2012Biochemistry}.  Since plectoneme hopping is related to nucleation, the large forces inhibiting the nucleation thus  repress the hopping.

We also found that plectoneme hopping is associated with significant energetic changes between the two phases.  We identify the nucleation  as the starting point of a hopping event. Then we collected numerous simulations (5000) that contain the hopping events and  aligned  the starting points of the hopping events together for averaging the energetics.   The  energy change of the S phase accumulates up to about $20k_BT$ before the nucleation, while that of the P phase falls down to about $20k_BT$. These significant energetic changes facilitate  the abrupt nucleating-vanishing processes (Fig. \ref{fig:4:i}).    After the formation of the new plectoneme by nucleation, the energetic changes rapidly converge to $0$.  The energetic offsets between the two phases thus makes the total energy change smaller as thermal fluctuations, i.e.,$1k_BT$(inset of Fig. \ref{fig:4:i}).  It should be noted that there is no persistent energetic injection into  the system. Thus the  energy change must be compensated by the complementary reversed processes (one of the multiple plectonemes vanishes, corresponding to reversed buckling transition shown in Fig. \ref{fig:4:i}).

  \section*{Methods}

  \begin{figure}[H]
  \centering
    \subfigure[]{
    \label{fig:5:a}
    \includegraphics[width=0.45\textwidth,angle=0]{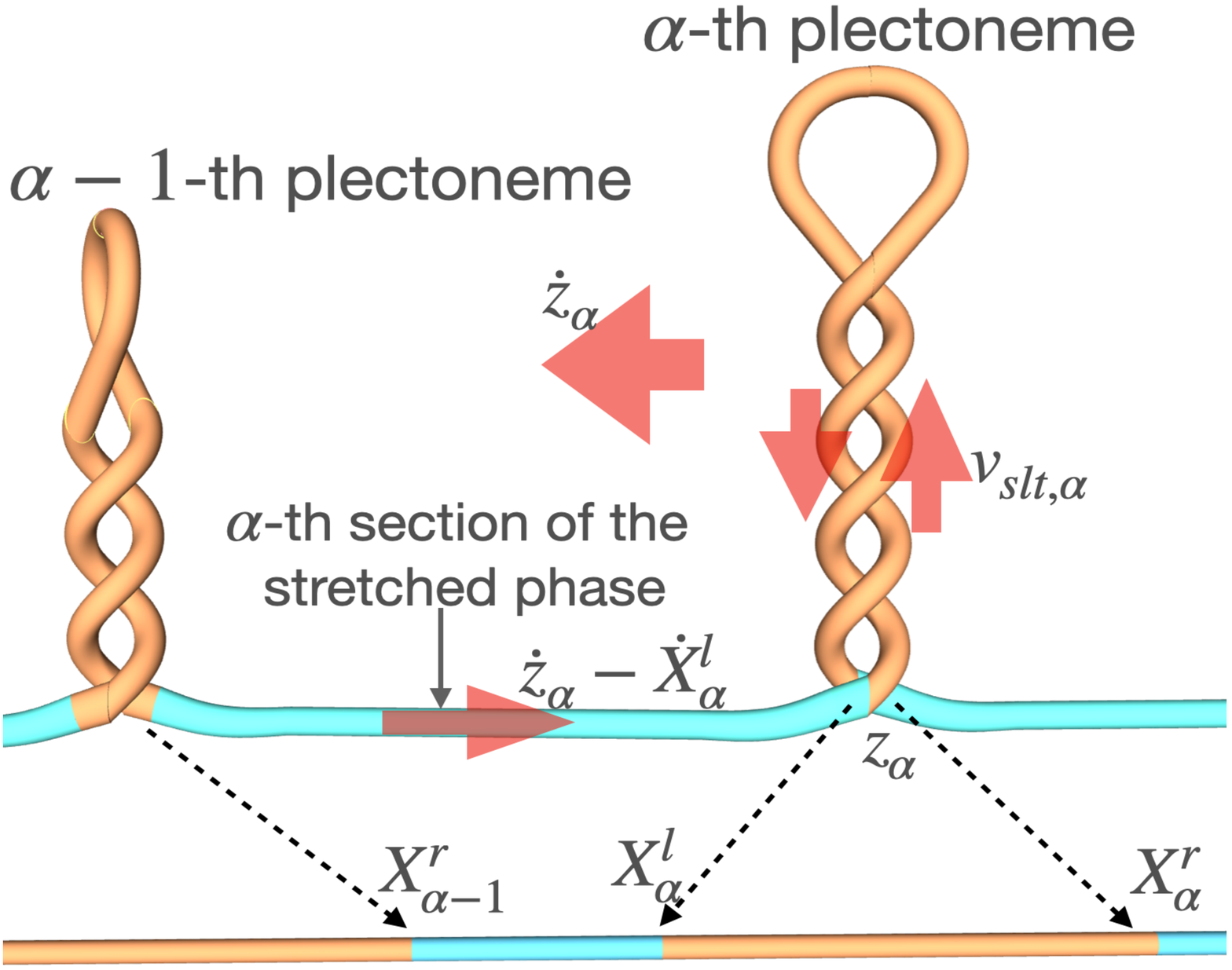}}
   \subfigure[]{
    \label{fig:5:b}
    \includegraphics[width=0.45\textwidth,angle=0]{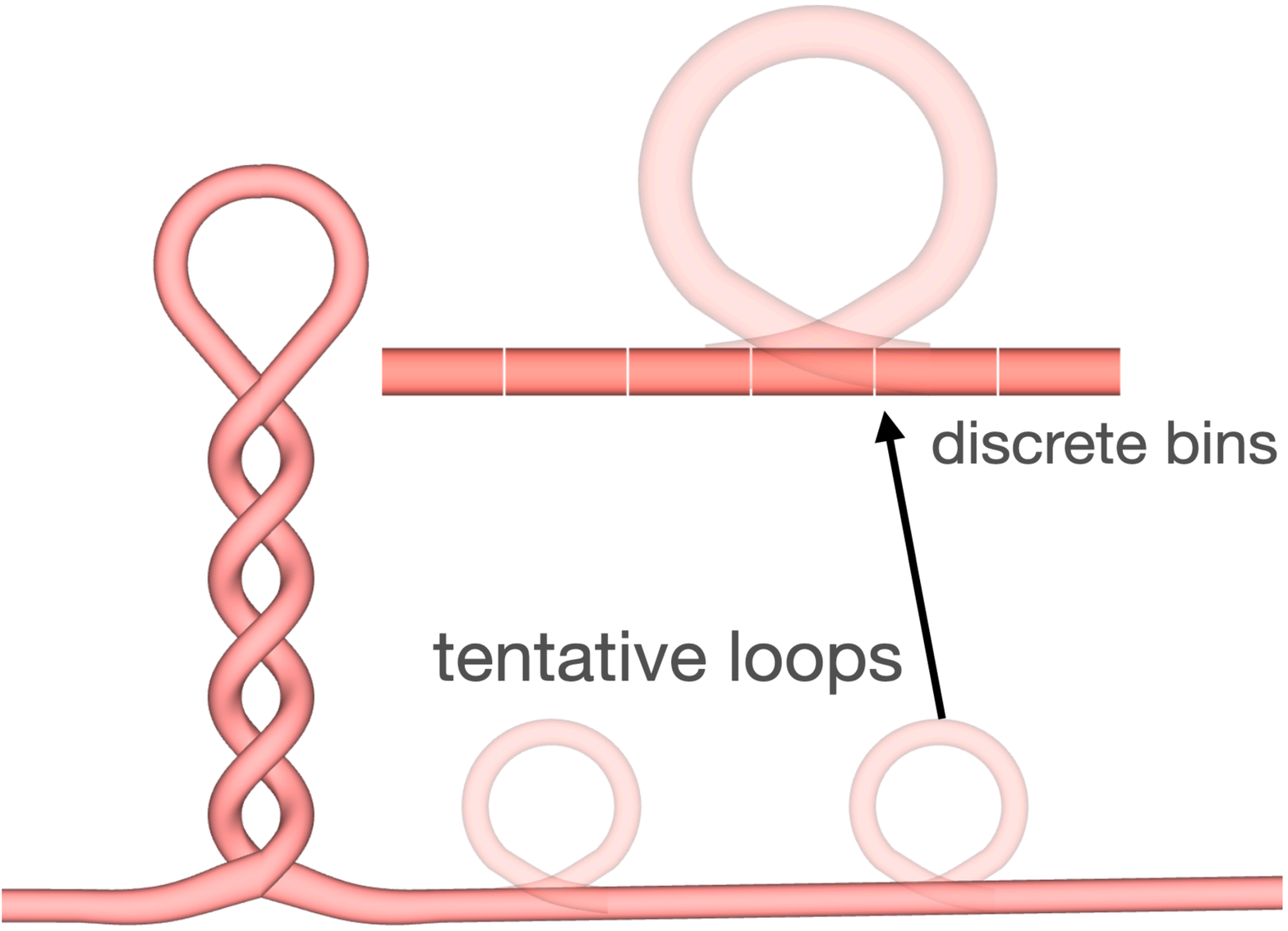}}
     \caption{\footnotesize The variables and tentative loops in the two-phase dynamics. (a) The phase boundaries $X_{\alpha}^{l/r}$ describe the phase-transformation between the stretched phase and the $\alpha$-th plectoneme, which involve  the transverse  displacement of the $\alpha$-th plectoneme described by $z_{\alpha}$, the displacement of  the $\alpha$-th  section of the stretched phase and the slithering motion of parallel segments of the plectoneme described by $v_{slt,\alpha}$. (b) The algorithm and the integrations are conducted in the discrete bins along the DNA axes. The tentative loops occupy  the bins.}
   \label{fig:5}
\end{figure}

\subsection*{The DNA topology and in-extensibility constraints}
 The Brownian dynamics Eq.\ref{eq:b2} is completed with addition of two constraints. The first one is the topology of supercoiling and the second one is the in-extensibility of the model DNA.   
 The DNA topology constraint, i.e., $\Delta Lk=const.$, gives the allocation,  
  \begin{equation}
  \Delta Lk = \Delta Lk_S + \Delta Lk_P,\tag{M1}
  \label{eq:m4}
     \end{equation}
where   $\Delta Lk_S$ and  $\Delta Lk_P$ are the excess linking numbers of  S and P phases, respectively. They can be written as the functions of $X$ and $\sigma_{S/P}^{eq}$, i.e., $\Delta Lk_S=\sum_{\alpha=0}\frac{\Omega_0}{2\pi}\int_{X_{\alpha}^{r}}^{X_{\alpha+1}^{l}}\sigma_{S}^{eq}(u,\mathbf{ X})du$ and $\Delta Lk_P=\sum_{\alpha=0}\frac{\Omega_0}{2\pi}\int_{X_{\alpha+1}^{l}}^{X_{\alpha+1}^{r}}\sigma_{P}^{eq}(\mathbf{ X})du$, where $\sigma_{S}^{eq}$ and $\sigma_{P}^{eq}$ satisfy  the torque equilibrium $c_S\sigma_{S}^{eq}(u,\mathbf{ X})=c_P\sigma_P^{eq}(X)=T_q^{in}$.

 The second constraint originates from the in-extensibility of the model DNA. Eq \ref{eq:b2} describes the phase-transformation motion perpendicular to the stretching direction.  As shown in Fig. \ref{fig:5:a},  $X_{\alpha}^{l,r}$ specify the $\alpha$-th plectoneme,  and also involve the displacements of  $\alpha$-th and $\alpha+1$-th sections of the stretched phase. This connectivity between the plectonemic and stretched phases additionally confines  the translational displacements parallel to the stretching direction. Plectonemes and stretched phase are subjected to the tensions at boundaries $f_u|_{u=X_{\alpha}^{l/r}}$, on which  potential $\Phi_0(\mathbf{ X})$ depends.   The components of motion parallel to the stretching direction,  involve the linear combinations of $X_{\alpha}^{l/r}$. Notice that the coordinate of the $\alpha$-th plectoneme is $z_{\alpha}=X_{\alpha}^{l}-\sum_1^{\alpha-1}l_{plect,\alpha’}$, where $l_{plect,\alpha’}=X_{\alpha’}^{r}-X_{\alpha’}^{l}$ is the contour length of the $\alpha’$-th plectoneme. The velocity of the $\alpha$-th stretched section is defined as $\dot{z}_{\alpha}-\dot{X}_{\alpha}^{l}$. The internal slithering velocity of the parallel segments of the plectoneme $v_{slt,\alpha}=\frac{1}{2}(\dot{X}_{\alpha}^l+\dot{X}_{\alpha}^r)$. The constraints on parallel components result in mechanical balances on the stretching direction
  
 \begin{equation}
    \begin{split}
&f_u|_{u=X_{\alpha}^{r}}-f_u|_{u=X_{\alpha}^{l}}-\mu_{slt}v_{slt,\alpha}=\gamma_{2,\alpha}\dot{z}_{\alpha}+\sqrt{2\beta^{-1}\gamma_{2,\alpha}}\dot{W}_{3}\\
&f_u|_{u=X_{\alpha}^{l}}-f_u|_{u=X_{\alpha-1}^{r}}=\gamma_{3,\alpha}(\dot{z}_{\alpha}-\dot{X}_{\alpha}^{l})+\sqrt{2\beta^{-1}\gamma_{3,\alpha}}\dot{W}_{4}\end{split},\tag{M2}
\label{eq:m5}
\end{equation}  
where $\mu_{slt}$ is slithering drag coefficient and  $\gamma_{2,\alpha}$  is the drag coefficient for the transverse displacement of the $\alpha$-th plectoneme, $\dot{W}_{3}$ is the white noises on the displacement, $\gamma_{3,\alpha}$ the drag coefficient for translational displacement of $\alpha$-th stretched section, $\dot{W}_{4}$ is the white noises on the motion of the $\alpha$-th stretched section. The first relation in Eq \ref{eq:m5} describes the constraints on the transverse  displacement of the  $\alpha$-th plectoneme, and the second one describes constraints between the $\alpha-1$ and $\alpha$-th plectonemes.    In the dWLC method, the effective viscous drags  is $6\pi\eta R_{H}$ for each segment. Thus $\gamma_{1,\alpha}=6\pi\eta R_{H}\frac{l_{plect,\alpha}}{l_0}\times\frac{1}{2}=0.6\pi\eta l_{plect,\alpha}$ where we have used $\frac{R_H}{l_0}\simeq 0.2$ and $\frac{1}{2}$ means a half of plectoneme contour length in Eq \ref{eq:b2},  $\gamma_{2,\alpha}=1.2\pi\eta l_{plect,\alpha}$ and $\gamma_{3,\alpha}=1.2\pi\eta(X_{\alpha}^{l}-X_{\alpha-1}^{r})$. 

Neverthless, the friction coefficient $\mu_{slt}$ for slithering is unknown. According to a relation of the slithering drag,  $\mu_{slt}\propto\frac{l_{plect,\alpha}}{ln(\frac{R_{plect}}{R})}$\cite{marko1994fluctuations}, where the plectoneme radius $R_{plect}\approx\chi_1ln(\frac{\chi_2}{f})$ and $\chi_1= 0.4 nm$ and $\chi_{2}=360 pN$\cite{van2012Dynamics}. Calibration of the two-phase model with the dWLC method gives $\mu_{slt}=\frac{0.4 l_{plect,\alpha}}{ln(\frac{R_{plect}}{R})}$.
For supercoiling with a single plectoneme, on average, $\dot{z}=v_{slt}$, and then the viscous drag is $\gamma_{1,1}+\gamma_{2,1}+\mu_{slt,1}$. According to the Einstein relation, the diffusion coefficient of individual plectoneme is  $D=\frac{k_BT}{\gamma_{1,1}+\gamma_{2,1}+\mu_{slt,1}}$.

 \subsection*{The routine for simulating the two-phase dynamics}

$i$. Generating the conformation $\{X_{\alpha}^{l/r},\sigma_{S/P}^{eq}\}$. 

The model DNA is discretized  as bins with a size $5nm$, on which the integrations are numerically conducted (Fig. \ref{fig:5:b}).  
The observable  $\mathbf{ X}$ follows Eq \ref{eq:b2} under constraints Eqs \ref{eq:m4} and \ref{eq:m5}.  We select the time-step as $dt=2.5\frac{\zeta_RL^2}{k_BTl_t}$, specifically, $\Delta t=0.1, 0.2, 0.4, 0.9$ and $5 ms$ for $3, 4.5, 6, 9$ and $21 kb$, respectively. At each time-step,  Eq \ref{eq:b2} and constraints Eqs \ref{eq:m4} and \ref{eq:m5}  together constitute the self-consistent equations that can be numerically solved with iteration procedure. Take 20 iteration cycles at each time-step, and the convergence (relative deviation of $\sigma_{S/P} < 0.001$) is achieved within 10 iteration cycles.    

$ii$.  Sampling new plectonemes.  The energy of the supercoiled DNA with n tentative loops for nucleation is  
  \begin{equation}
\Phi_{n}(\mathbf{ X})=\Phi_0(\mathbf{ X})+\sum_{i}^{n}E_{loop,i}, \tag{M3}
\label{eq:m6}
\end{equation}
where $E_{loop,i}=8k_BT\sqrt{\frac{l_b f_u}{k_BT}}$  is  the energy penalty for looping at location $u$\cite{Marko2012Competition}. The excess linking number of the $n$ tentative loops is  $\Delta Lk_l=n+\frac{1}{2\pi}\frac{\Delta\theta}{\Delta u}\sum L_{loop,i}$, where the relative twist is proportional to torque, i.e., $\frac{\Delta\theta}{\Delta u}=\frac{T_q^{in}}{k_BTl_t}$. The time-scale for looping is on microseconds\cite{Daniels2011Nucleation}, much shorter than $dt$. We thus assume that a portion of  $\Delta Lk_S$ is stored in loops, then the tentative state follows the two constraints $L_S=L_S’+\sum L_{loop,i}$ and $\Delta Lk_S=\Delta Lk_S’+\Delta Lk_l$ where $L_S’$ and $Lk_S’$ are the contour length and linking number of  the S phase of the tentative state, respectively, and the contour length of a loop is $L_{loop,i}=\sqrt{2\pi^2}\sqrt{k_BTl_b/f_u}$  \cite{Marko2012Competition}.

  \begin{figure}[H]
  \centering
    \label{fig:6}
    \includegraphics[width=0.8\textwidth,angle=0]{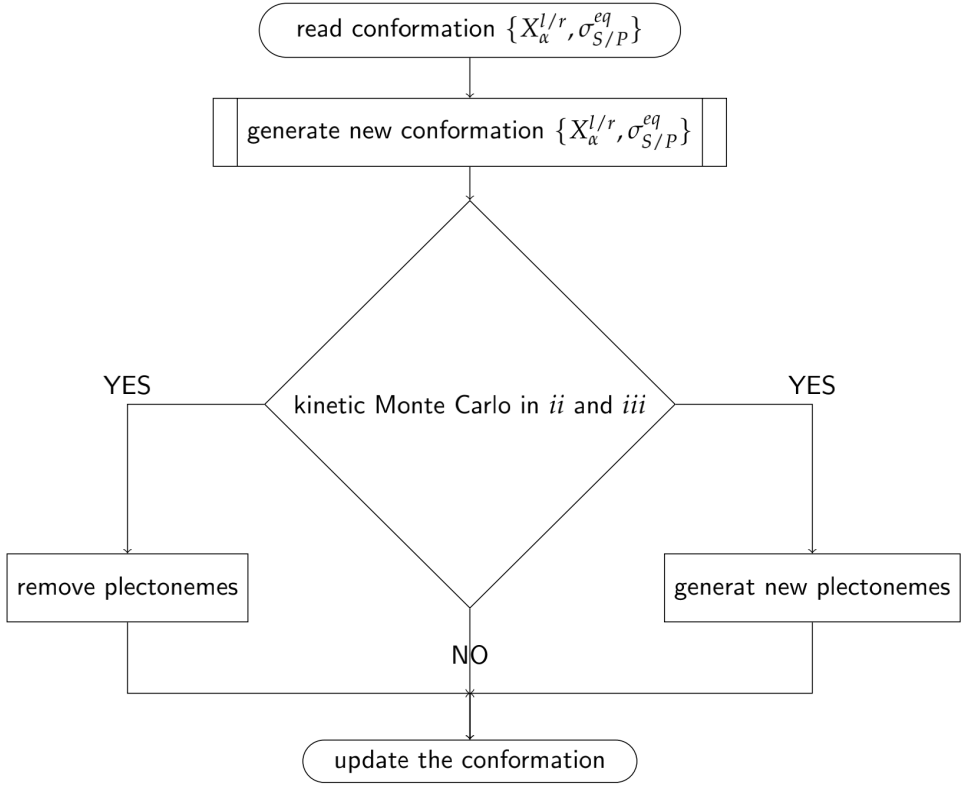}
     \caption{\footnotesize Flow chart of simulation routine.}
   \label{fig:6}
\end{figure}

  The loops along the stretched DNA can be regarded as hard-core particles  occupying a 1-dimensional discrete bins (Fig. \ref{fig:5:b}), where each bin characterizes the size of the thermal  fluctuations of the DNA.    The length of each bin $d_{bin}\approx 5 nm$, defined as the transverse amplitude of the fluctuations\cite{Marko1995Statistical,Marko2012Competition}.  Define a partition function of the tentative state with n loops,

 \begin{equation}
 \mathcal{P}_n=\sum_{I_1<I_2<\cdot\cdot\cdot<I_n} exp[-\beta\Phi_n(\mathbf{X})],\tag{M4}
 \label{eq:m7}
  \end{equation}
  where $I_i$ denotes the location of the i-th tentative loop. Accordingly, $P_0=\frac{\mathcal{P}_0}{\sum_i\mathcal{P}_i}$ is the probability of the supercoiling without tentative loops and $P_n=\frac{\mathcal{P}_n}{\sum_i\mathcal{P}_i}$ 
 is the probability of supercoiling with $n$ tentative loops.  According to the kinetic Monte Carlo method\cite{2009Chapter}, generate a random $\mathcal{R}$ in (0,1). The range (0,1) is divided into $(0,P_0]$, $(P_0,P_0+P_1]$,$\cdot\cdot\cdot$,$(\sum_i^{\mathcal{N}-1}P_i,\sum_i^{\mathcal{N}}P_i],\cdot\cdot\cdot$, where $\sum_{\mathcal{N}+1}^{\infty}P_i\simeq 0$  and  $\sum_i^{\mathcal{N}}P_i\simeq 1$ (often $\mathcal{N}\leqslant 2$). If $\mathcal{R}$ lies in $(\sum_i^{N-1}P_i,\sum_i^{N}P_i]$, the state with n loops is sampled.  The same procedure is applied to locate the $n$ loops according to the distribution $\frac{exp[-\beta\Phi_n(\mathbf{X})]}{\mathcal{P}_n}$. 
   (For a  fully stretched DNA, distribution Eq.\ref{eq:m7} implies an entropy brought by $n$ loops, i.e., $S_n=k_Bn ln(\frac{L-n L_{loop}}{d_{bin}})-k_Bln(n!)$, the $\frac{L-n L_{loop}}{d_{bin}}$ bins occupied by $n$ identical particles \cite{Marko2012Competition}.)
     
$iii$. Removing plectonemes
 
   If $\mathcal{N}’$ plectonemes satisfy $l_{plect,\alpha}<\sqrt{2\pi^2}\sqrt{k_BTf/l_b}$ (often $\mathcal{N}’\leqslant2$). Similar to $ii$, remove the plectonemes according to the distribution $exp[-\beta\Phi_n(\mathbf{X})]$.
$ii$ and $iii$  generate and remove plectonemes according to the Boltzmann distribution.

Update the conformation $\{X_{\alpha}^{l/r},\sigma_{S/P}^{eq}\}$, and then repeat $i$, $ii$ and $iii$. All the steps can be organized as the  flow chart (Fig. \ref{fig:6}).

 \section*{Discussion and Conclusion}

In this study, based on the polymer physics or the WLC model of DNA, we have constructed a two-phase dynamic model of the DNA supercoiling. To establish such a model,  we demonstrate the fast dynamics of torque equilibrium within the S and P (plectonemic) phases,  identify the phase boundaries as slow observables, and  derive the energetics associated with the slow observables. By comparing with the WLC model, we have shown that the two-phase dynamic model provides a physically justified and efficient way in representing DNA supercoiling dynamics. The model is particularly suited for studying the plectoneme dynamics from milliseconds to seconds. Correspondingly, we have demonstrated the plectoneme diffusion, nucleation,  and hopping, consistently with measurements from single molecule experiments\cite{2017Single,Sheinin2012Biochemistry,van2012Dynamics}.  

  In order to probe the time-scale separation, we have  investigated the dynamics of torque approaching to equilibrium within the pure S and P phases, respectively.    Based on the WLC model,  the internal torque transports as  twist angle propagates. Similarly,  by introducing an effective twist angle of the plectonemic helix, the internal torque propagation within the P phase can be described, comparably.   Such derivations are confirmed by numerical simulations based on the dWLC, which consistently show that the torque transport is fast, i.e., on either the S or P phase of the supercoiled DNA.    

Accordingly, we choose the  the boundaries of the S and P phases as observables to describe the slow dynamics.   Based on the interphase torque equilibrium, one can then derive the free energy associated with the collective phase boundaries, and the corresponding equation of motion via the Langevin dynamics.

To compare torsional responses from the two-phase model with that from the WLC model, we tested at a constant rate of supercoil accumulation, $\Delta Lk=10$ per second, comparable to the RNA polymerization rate during transcription elongation (i.e.,$\sim$ 100 bp per sec), in which $\Delta Lk=0.1$ is injected into DNA for the RNA polymerase enzyme unwinding to synthesize one nucleotide\cite{thomen2005unravelling,Jing2018How}. The discontinuities on torque and extension curves are reproduced by the two-phase dynamic model, which reflect the buckling transition during plectoneme nucleation\cite{Forth2008Abrupt,Marko2012Competition,Emanuel2013Multiplectoneme,Daniels2011Nucleation,Walker2018Dynamics,ott2020dynamics}. The torque and extension versus linking number curves  suggest that the non-equilibrium effect of the supercoiling accumulation (from the growth of the plectonemic phase) induced by RNAP is constantly  relaxed during each polymerization  step or elongation cycle.  Hence, such an RNA elongation process can be regarded as quasi-steady state.    An obvious advantage of the currently developed two-phase dynamic model of DNA supercoiling comes from its trivial computational cost as we only deal with the slow degrees of freedom of the system.     Besides, we  performed simulations of the plectonemes growth under a constant torque since some studies also regarded RNAPs as a torsional motor which imposes a constant torque\cite{Mielke2004Transcription}.  We accordingly determined  frictional coefficient due to slithering in plectoneme by calibrating with the WLC model.

With the two-phase dynamic model we described the DNA supercoiling dynamics and particularly reproduced the plectoneme diffusion, nucleation and hopping.  The plectoneme diffusion coefficient is about 0.1 to 0.2$\mu m^2/s$,  decreasing slightly with the  stretching force. Large force indeed makes the plectoneme more compact, which increases the friction for the slithering of parallel segments of the plecoenemes. Consequently, the large force reduces the diffusive capability of plectoneme.    Large stretching force also suppresses the nucleation rate, as the force increases the energy penalty for forming a loop. The ability of salt concentration to screen the electrostatic repulsion between opposing segments in plectonemes also affects the radius of plectonemic helix\cite{1997The,van2012Dynamics}. Similar to large stretching force, strong ionic strength results in compact plectonemes. The resulting  compact plectonemes require higher energy penalty to fluctuate, i.e., large stretching force or strong ionic strength can thus stabilize plectonemes. Accordingly, large force or strong ionic strength can prolong the mean life-time of the individual plectoneme. We also observed that a plectoneme can perform  a long distance hopping. Specifically, the abruptly conformational change is accompanied by significant energetic changes.  However, the total energetic change of the two phase  is trivial or  as small as thermal fluctuations. 

Recent experiments reveal the long-distance cooperative and antagonistic dynamics between multiple RNAPs\cite{kim2019long}. The experimental evidences suggest that the superpcoil cancellation between adjacent RNAPs makes similar apparent transcription elongation rate. However, upon promoter repression or divergently transcribed genes, RNAPs switch their behavior from collaborative to antagonistic due to the build-up supercoils.   There are two important dynamic behaviors that are incorporated in our two-phase model.  First,  RNAPs can be sensitive to the torque close to the stalling condition (5$\sim$11 $pN$ $nm$)\cite{Ma2013Transcription,2019Transcription}.  A static  coexistence torque may not slow down or even stall the transcription elongation, but the torque  fluctuation (time-dependent) may.  Thus time-dependent torque may induce effects on the long-distance cooperative behaviors of RNAPs.
Second,  supercoil can fast propagate as twist (torque transport) or slowly propagate as writhe (e.g, plectoneme diffusion). These two kinds of supercoiling propagations are often confounded  in modeling supercoiling-dependent dynamics.  During transcription elongation, the build-up supercoil can be stored in writhe and can be transmitted as twist that can be felt by RNAPs as the torsional stress\cite{2019Transcription,Ma2014RNA,Ma2014Interplay,Ma2013Transcription}.   The two-phase dynamic model of DNA supercoiling is expected to be applied to improve understanding the supercoiling-dependent transcription.

DNA supercoiling facilitates the site juxtaposition of two distal  DNA sites  by slithering opposing segments of the interwound superhelix or bending DNA to coil\cite{oram1997communications,polikanov2007probability,Parker1991Dynamics,Marko1997The}, it then promotes protein binding or enhance the gene expression\cite{2008How,liu2001dna,2009Facilitated}.
  As we have seen, slithering motion of  the segments in a plectomeic helix is accompanied with plectoneme diffusion.   However, the diffusion may not offer an efficient mechanism  for site juxtaposition. An efficient mechanism for achieving  site juxtaposition  is combining plectoneme diffusion and hopping.   These mechanisms can be efficiently  explored by employing the two phase dynamics of the DNA supercoiling which is physical justified and computationally trivial.

 \section*{Acknowledgements}

This work has been supported by NSFC Grant $\#11775016$ and $\#11635002$. JY has been supported by the CMCF of UCI via NSF DMS 1763272 and the Simons 
Foundation grant $\#594598$ and start-up fund from UCI. We acknowledge the computational support from the Beijing Computational Science Research 
Center (CSRC). Thanks Dr Xinliang Xu for discussions on the work.

%

\end{document}